\documentclass[prr,longbibliography,amsfonts,floatfix,superscriptaddress,twocolumn, aps]{revtex4-1}

\usepackage{mathbbol}
\usepackage{amsmath,amssymb,bm,amsthm}
\usepackage{braket}
\usepackage{soul}
\usepackage{dsfont}
\usepackage{gensymb}
\usepackage{times}
\usepackage{multirow}
\usepackage[utf8x]{inputenc}
\usepackage[usenames,dvipsnames, pdftex]{xcolor}
\usepackage[colorlinks,bookmarks=false,citecolor=blue,linkcolor=red,urlcolor=blue]{hyperref}
\usepackage{graphicx}
\usepackage[makeroom]{cancel}
\usepackage[normalem]{ulem}
\usepackage[caption=false]{subfig}		
\usepackage{floatrow}
\usepackage{pdfpages}
\usepackage{pgffor}

\makeatletter
\AtBeginDocument{\let\LS@rot\@undefined}
\makeatother

\newcommand{\LCO}{${\mathrm{La}}_{2}{\mathrm{CuO}}_{4}$}
\newcommand{\CCO}{$\mathrm{CaCuO}_2$}
\newcommand{\SIO}{$\mathrm{Sr}_2\mathrm{IrO}_4$}

\begin{document}

\title{Magnon Spectra of Cuprates beyond Spin Wave Theory}


\author{Jiahui Bao}
\email{jiahui.bao@oist.jp}
\affiliation{Theory of Quantum Matter Unit, Okinawa Institute of Science and Technology Graduate University, Onna-son, Okinawa 904-0495, Japan}

\author{Matthias Gohlke}
\affiliation{Theory of Quantum Matter Unit, Okinawa Institute of Science and Technology Graduate University, Onna-son, Okinawa 904-0495, Japan}

\author{Jeffrey G. Rau}
\affiliation{Department of Physics, University of Windsor, 401 Sunset Avenue, Windsor, Ontario, N9B 3P4, Canada}

\author{Nic Shannon}
\affiliation{Theory of Quantum Matter Unit, Okinawa Institute of Science and Technology Graduate University, Onna-son, Okinawa 904-0495, Japan}


\begin{abstract}
The usual starting point for understanding magnons in cuprate antiferromagnets such as \LCO\  
is a spin model incorporating cyclic exchange, which descends from a 
one-band Hubbard model, and has parameters taken from fits based 
on non-interacting spin wave theory. 
Here we explore whether this provides a reliable description of experiment, 
using matrix product states (MPS) to calculate      
magnon spectra beyond spin wave theory.
We find that analysis based on low orders of spin wave theory leads to 
systematic overestimates of exchange parameters, with corresponding 
errors in estimates of Hubbard $t/U$.
Once these are corrected, the ``standard'' model provides a good 
account of magnon dispersion and lineshape in \LCO, 
but fails to fully capture the continuum observed at high energies.  
The extension of this analysis to \CCO\ and \SIO\ is also discussed.   
\end{abstract}

\maketitle

\textit{Introduction.} 
Cuprate perovskites, typified by \LCO, are an important class of magnetic insulators,  
providing some of the best examples of quasi-two-dimensional antiferromagnets \cite{Chakravarty1989}, 
as well as being parent materials for high--temperature superconductors \cite{Bednorz1986,Anderson1987}.  
Early attempts to understand magnon spectra in \LCO\ rested on the 
antiferromagnetic (AF) Heisenberg model on a square lattice \cite{Hayden1991}, 
as derived from a one--band Hubbard model in the limit \mbox{$t/U \to 0$}~\cite{Gros1987}. 
More detailed measurements of magnon dispersion at high energy revealed a dispersion on 
the magnetic Brillouin zone boundary which could not be understood in terms of a simple 
Heisenberg AF \cite{Coldea2001}.
This lead to the adoption of a more complicated model, including four-spin cyclic exchange \cite{Coldea2001,Katanin2002,Toader2005,Headings2010,Dean2012,Yamamoto2019,Robarts2021}, 
derived from the same one-band Hubbard model at finite $t/U$  
\cite{Takahashi1977,MacDonald1988,Roger1989,MacDonald1990,Roger1989}.  
Variants of this model have been also used to analyze magnon spectra in 
\CCO\ \cite{Martinelli2022,Peng2017}, 
$\mathrm{SrCuO}_2$~\cite{Wang2024}, 
and \SIO\ \cite{Bertinshaw2019,Lu2020,Kim2012,Vale2015,Pincini2017}.   

To date, where such models have been compared with experiment, the analysis has mostly rested 
on linear spin wave theory (LSWT) \cite{Coldea2001,Toader2005,Headings2010,Dean2012,Yamamoto2019,Robarts2021}, 
or LSWT with leading $1/S$ corrections \cite{Katanin2002}.
However, 
low-order spin-wave approximations 
can show significant deviations from quantum calculations \cite{Chi2022}.
In related work on the Heisenberg AF 
\cite{Singh1995,Syljuasen2000,Sandvik2001,Luescher2009,Powalski2015,Shao2017,Masaki-Kato-unpub,Verresen2018}, 
and its material instantiation 
\mbox{${\mathrm{Cu(DCOO)}}_2 \cdot \mathrm{4D}_2\mathrm{O}$} (CFTD)~\mbox{\cite{Ronnow2001,Chritensen2007,Piazza2014}}, 
quantum effects missing from these theories were found to have a significant impact 
on zone-boundary magnons.
Moreover, estimates of Hubbard $t/U$ taken from spin wave fits 
in \LCO\ \cite{Coldea2001,Headings2010,Toader2005,Katanin2002,Yamamoto2019} 
and \CCO\ \cite{Martinelli2022,Peng2017} are consistently higher than {\it ab initio} values \cite{Hirayama2018,Hirayama2019,Schmid2023}.  
For this reason, it is important to understand whether the model currently used to understand magnetism 
in \LCO\ provides a reliable description of experiment.
This question is particularly significant in the light of the ongoing efforts 
to understand superconductivity in doped cuprates \cite{Kowalski2021}.
Magnon spectra can provide important information about electronic interactions~\cite{Coldea2001}.   
But to access this, it is necessary to disentangle effects arising from the spin-wave approximation.

\begin{figure*}
\centering
	\includegraphics[height=0.25\textheight]{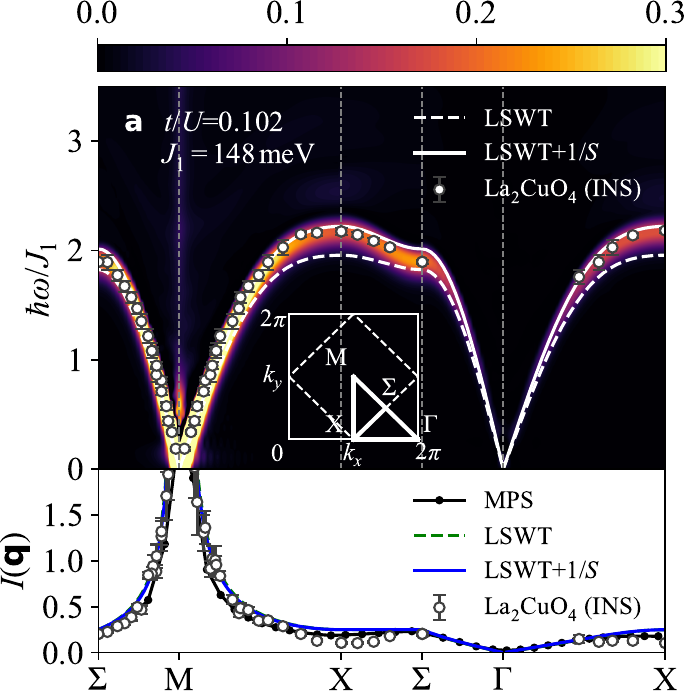}
	\hspace{-0.05cm}
	\includegraphics[height=0.25\textheight]{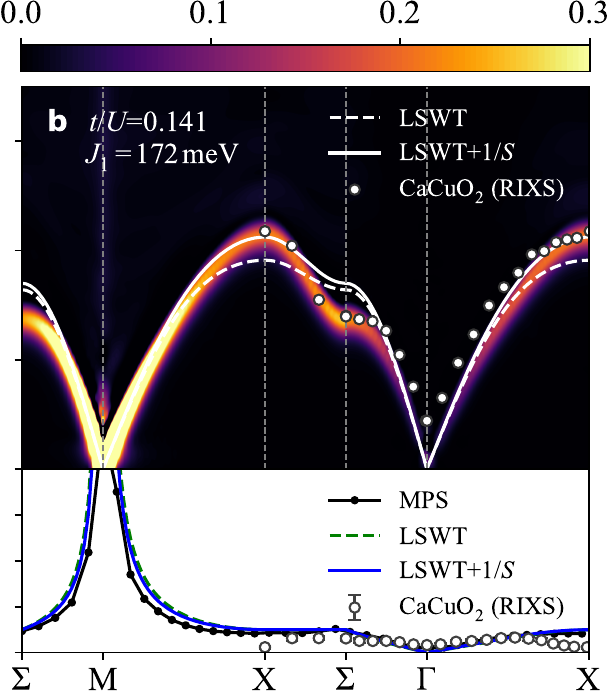}
	\hspace{-0.05cm}
	\includegraphics[height=0.25\textheight]{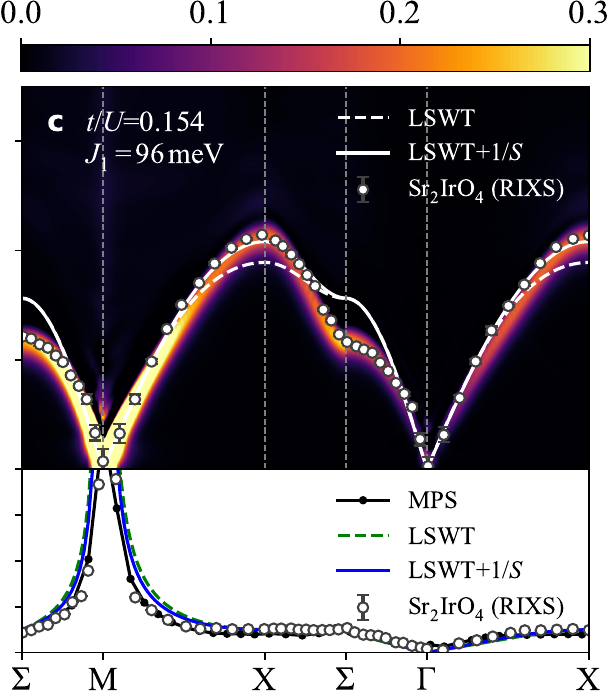}
	\vspace{-0.1cm}
	\\
	\captionsetup[subfigure]{labelformat=empty}
	\sidesubfloat[]{\includegraphics[width=0.405\textwidth]{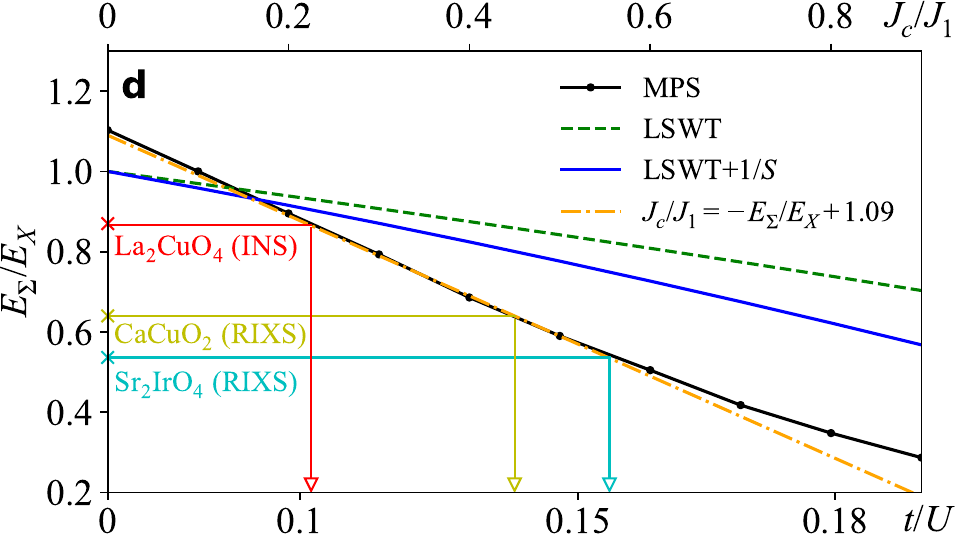}}
	\hspace{-0.3cm}
	\sidesubfloat[Estimates of parameters for ${\mathrm{La}}_{2}{\mathrm{CuO}}_{4}$  \label{fig:parameters}]{
		\renewcommand\arraystretch{1.13}
		\begin{tabular}[t]{| l | l | l | l | l | l |}
		\hline
		$\bold{e}$&$J_1$[meV]&$J_2$[meV]&$J_3$[meV]&$J_c$[meV]&$t/U$\\
		\hline
		INS (low\ T)~\cite{Coldea2001} & $146.3\pm4$ &$3.1\pm0.4$&$3.1\pm0.4$&$61\pm8$&0.136\\
		INS (low T)~\cite{Headings2010} & $143\pm2$ &$2.9\pm0.2$&$2.9\pm0.2$&$58\pm4$&0.134\\
		INS (high\ T)~\cite{Toader2005} &139.8&2.6&2.6 &55.9&0.108\\
		sc-SWT~\cite{Katanin2002} & 151.9 & 3.8 & 3.8 & 36.5 & n/a \\
		m-SWT~\cite{Yamamoto2019}&145&1.6&1.6&32&0.102\\
		LSWT+1/$S$ (*) &148 & 2.3 & 2.3 & 45.9 & 0.120 \\
		MPS (*) & 148 & 1.6 & 1.6 & 32.6 & 0.102 \\
		\hline
		\end{tabular}}
\caption{
Magnon spectra of cuprate and iridate magnets, compared with predictions descended from a one-band Hubbard model.
(a)~Dispersion and intensity of excitations in ${\mathrm{La}}_{2}{\mathrm{CuO}}_{4}$, as found in calculations of 
$S_\perp({\bf q},\omega)$ for the effective spin model ${\mathcal H}_\sigma$ 
[Eq.~(\ref{eq:spin.model})], using matrix product states (MPS), linear spin wave theory (LSWT), and
interacting spin wave theory \mbox{(LSWT + 1/S)}.
Points show the results of inelastic neutron scattering (INS) experiments~\cite{Headings2010}, 
with energy measured in units of the first-neighbor interaction $J_1$. 
(b)~Equivalent results for $\mathrm{CaCuO}_2$, compared with results of resonant inelastic x-ray scattering (RIXS)~\cite{Martinelli2022}.
(c)~Results for $\mathrm{Sr}_2\mathrm{IrO}_4$, characterized using RIXS~\cite{Kim2012}.
(d)~Ratio of excitation energies at ${\bf q}_\Sigma =(\pi/2,\pi/2)$ and ${\bf q}_X=(\pi, 0)$, 
showing scaling with $J_c/J_1 \sim t^2/U^2$ [Eq.~(\ref{eq:emp.relation})].
(e)~Parameters for ${\mathrm{La}}_{2}{\mathrm{CuO}}_{4}$ reported in the experimental 
literature \cite{Coldea2001,Headings2010,Toader2005}, and subsequent analysis using self-consistent 
spin wave theory (sc-SWT)~\cite{Katanin2002} and modified spin wave theory (m-SWT)~\cite{Yamamoto2019}.
Results from this study are indicated with (*). 
Where fits have been characterized using the one-band Hubbard model, 
values are quoted for $t/U$.   
}
\label{fig:1}
\end{figure*}

The goal of this Letter is to explore how well the model commonly used to describe magnetic excitations
in cuprate antiferromagnets \cite{Coldea2001,Katanin2002,Toader2005,Headings2010,Dean2012,Yamamoto2019,Robarts2021}
fits experiment, once quantum effects beyond spin wave theory are taken into account.   
To this end, we use numerical simulation based on matrix product states 
(MPS)~\cite{White1992, Schollwock2011, Haegeman2011, Haegeman2016, Gohlke2017, Laurens2019, Hauschild2018} 
to explore the magnon spectra of cuprate materials.
We make explicit comparison with experimental results for 
${\mathrm{La}}_{2}{\mathrm{CuO}}_{4}$, 
obtaining fits to magnon dispersion and intensity, and estimating the corresponding value of 
$t/U$ within a one-band Hubbard model. 
We explore the errors which arise in analysis of experiment based on LSWT, 
and introduce a method of estimating $t/U$ directly from magnon energies at 
high-symmetry points. 
We find good agreement with magnon dispersion and lineshapes in ${\mathrm{La}}_{2}{\mathrm{CuO}}_{4}$, 
with the exception of  the high-energy continuum observed near ${\bf q}_X = (\pi,0)$ \cite{Headings2010,Robarts2021}.
The analysis of magnon dispersion is extended to $\mathrm{CaCuO}_2$ 
and $\mathrm{Sr}_2\mathrm{IrO}_4$, with consistent findings. 
Key results are summarized in Fig.~\ref{fig:1}.   
We draw two main conclusions:
(i) that fits to magnon spectra based on LSWT have lead to 
systematic errors in estimates of exchange parameters in  
cuprate AF's; and 
(ii) that the model provides a good description of  spin-wave excitations in 
cuprate AF's, but not of the continuum at high energies.

\begin{figure}[tp]
  \centering
  \includegraphics[width=1\textwidth]{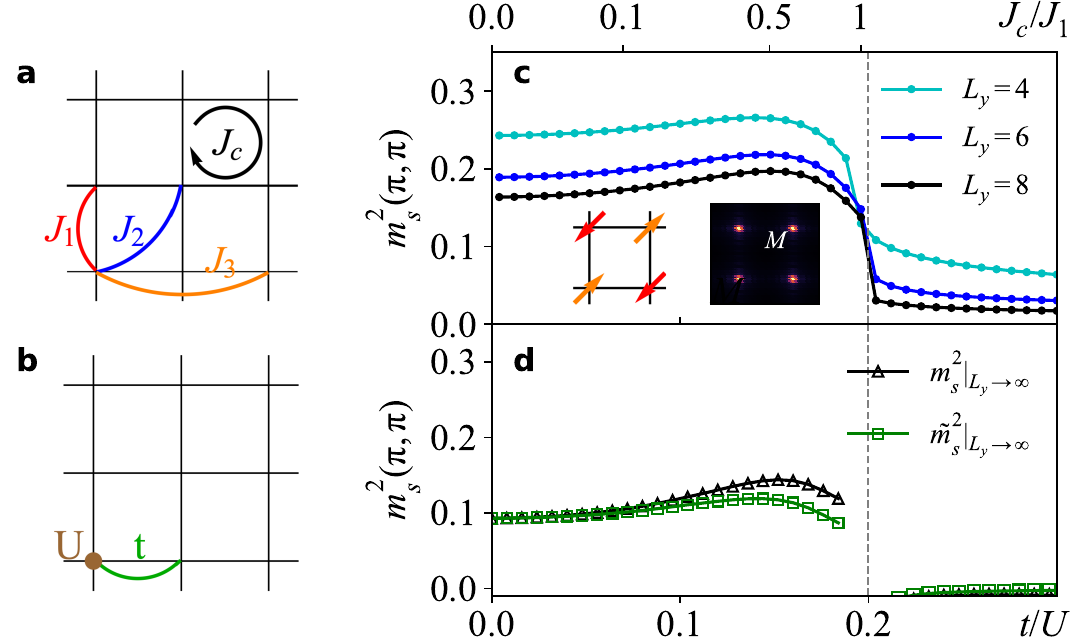}
  \caption{
  Interactions and ground-state properties of model for magnetism in cuprate antiferromagnets.
  (a) Parameters of effective spin model ${\mathcal H}_\sigma $ [Eq.~(\ref{eq:spin.model})].
  (b) Parameters of parent one-band Hubbard model ${\mathcal H}_{\sf U}$.
  (c) Staggered magnetization squared, $m_s^2(\pi,\pi)$, found in  
  calculations based on matrix product states (MPS), as a function of $t/U$, (and equivalently, $J_c/J_1$).   
  The corresponding N\'eel order is shown in an inset.
  (d) Equivalent results for $m_s^2(\pi,\pi)$ in thermodynamic limit, $L_y \to \infty$.    
  MPS results for the spin model are shown using triangles.
  Predictions $\tilde{m}_s^2(\pi,\pi)$, taking into account charge fluctuations \cite{Delannoy2005}, 
  are denoted with squares.  
  All calculations were carried out for ${\mathcal H}_\sigma$ [Eq.~(\ref{eq:spin.model})], on cylinders of circumference $L_y$.
  }
  \label{fig:2}
\end{figure}

\textit{Model}. 
%
The model we consider is the one commonly used to analyze magnon spectra 
in ${\mathrm{La}}_{2}{\mathrm{CuO}}_{4}$ \cite{Coldea2001,Katanin2002,Toader2005,Headings2010,Dean2012,Robarts2021},  
\begin{eqnarray}
{\mathcal H}_\sigma
	&= &J_1 \sum_{\langle ij \rangle_1} \bold{S}_i\cdot \bold{S}_j 
	+ J_2 \sum_{\langle ij \rangle_2} \bold{S}_i\cdot \bold{S}_{j}  
	+ J_3 \sum_{\langle ij \rangle_3} \bold{S}_i\cdot \bold{S}_{j} \nonumber\\
	&&+ J_c \sum_{\langle ijlk \rangle} \big{[}(\bold{S}_i\cdot \bold{S}_j )(\bold{S}_k\cdot \bold{S}_l) + (\bold{S}_j\cdot \bold{S}_k)(\bold{S}_l\cdot \bold{S}_i ) \nonumber\\
	&& -(\bold{S}_i\cdot \bold{S}_k)(\bold{S}_j\cdot \bold{S}_l)\big{]}  \; ,
\label{eq:spin.model}
\end{eqnarray}
where exchange interactions on first, second and third-neighbor bonds,
compete with four-spin terms originating in cyclic exchange \cite{Dirac1929,Thouless1965,Roger1989}.
The relevant interactions are illustrated in Fig.~\ref{fig:2}(a).


This spin model
can in turn be derived from the one-band Hubbard model which provides the minimal description 
of ${\mathrm{La}}_{2}{\mathrm{CuO}}_{4}$ \cite{Anderson1987} 
\begin{eqnarray}
{\mathcal H}_{\sf U} 
	=  -t \sum_{\langle ij \rangle_1 \sigma } \left[ c^\dagger_{i\sigma} c^{\phantom\dagger}_{j\sigma} + \text{H.c.} \right] 
	+ U \sum_i c^\dagger_{i\uparrow} c^{\phantom\dagger}_{i\uparrow} c^\dagger_{i\downarrow} c^{\phantom\dagger}_{i\downarrow} \; ,
\label{eq:Hubbard.model}
\end{eqnarray}
cf. Fig.~\ref{fig:2}(b).   
Within an expansion about half-filling \cite{Takahashi1977,MacDonald1988,Roger1989,MacDonald1990,Roger1989}, the 
corresponding exchange parameters are
	$J_1 = 4 t^2/U \left( 1- 6 t^2/U^2 \right),\,
	J_c=20J_2 = 20J_3  = 80 \left( t^4/U^3 \right).$
\noindent
Further details of this mapping are given~\cite{supplemental.material}.
In \LCO, {\it ab initio} estimates suggest that 
\mbox{$t/U \approx 0.1$ \cite{Hirayama2018,Hirayama2019}},  and the \mbox{four-spin} 
interaction $J_c$ is therefore a significant fraction of the first-neighbor
exchange $J_1$.
We use ${\mathcal H}_\sigma$~[Eq.~(\ref{eq:spin.model})], 
with parameters appropriate to the one-band Hubbard model ${\mathcal H}_{\sf U}$, as the basis for all of the calculations in this Letter.
Except where comparing to experimental data, we set $\hbar = 1$.

\textit{Methods}. 
%
Matrix product states~\cite{Schollwock2011} 
provide a means of calculating ground-state and dynamical properties of 
quantum systems, which is equally capable of describing the magnons 
associated with conventional magnetic order~\cite{Verresen2018}, 
and the fractionalized excitations found in quantum spin chains \cite{Lake2013} 
and quantum spin liquids \cite{Gohlke2017}.
The calculations in this Letter were carried out  for a square lattice wrapped 
onto a cylinder, and  are not subject to any intrinsic bias, but are subject to 
corrections coming from from finite cylinder circumference and MPS bond dimension.
Technical details, including an analysis of convergence, 
are provided in the Supplemental Material~\cite{supplemental.material}.

\textit{Ground states.}
%
We first use density matrix renormalization group (DMRG)~\cite{White1992,Phien2012} 
to determine the ground state of the effective spin model [Eq.~(\ref{eq:spin.model})].
The N\'eel order found in ${\mathrm{La}}_{2}{\mathrm{CuO}}_{4}$ has ordering vector $\bold{q}_M=(\pi,\pi)$, 
and is characterized by a staggered magnetization  
\begin{equation}
	m^2_s(\pi,\pi) 
		= \frac{1}{N^2} \sum_{ij} e^{i(\pi,\pi) \cdot(\bold{r}_i-\bold{r}_j)} 
			\langle \bold{S}_i \cdot \bold{S}_j \rangle \; .
\label{eq:mS}
\end{equation}
In Fig.~\ref{fig:2}(c) we show results for $m^2_s(\pi,\pi)$ as a function of $t/U$ for infinite 
cylinders with circumference  $L_y = 4, 6, 8$.
Interpolating $L_y \to \infty$ by
	$m_s^2(\bold{q})|_L=m_s^2(\bold{q})|_\infty+a/L+b/L^2 + {\mathcal O}(1/L^3)$ (cf., e.g., \cite{Jiang2012}),
we find that $m_s(\pi,\pi)$ takes on a finite value 
for $t/U \lesssim 0.2$, and vanishes for larger values of $t/U$ [Fig.~\ref{fig:2}(d)], 
consistent with published results~\cite{Larsen2019}.
Where comparing with experiment, it is also necessary to take into account charge fluctuations 
at ${\mathcal O}(t^4/U^3)$ \cite{Delannoy2005}, which turns the magnetization into
$\tilde{m}_s = (1-2zt^2/U^2) m_s$, where $z=4$ is the coordination number.
Allowing for these, the staggered magnetization 
for  \mbox{$t/U \lesssim 0.15$} is largely independent of $t/U$, and takes 
on a value of $\tilde{m}_s(\pi,\pi) \approx 0.3$ [Fig.~\ref{fig:2}(d)].
Assuming isotropic $g \approx 2$, this translates into an ordered moment of $\sim 0.6\ \mu_{\sf B}$, 
consistent with earlier theory for the Heisenberg AF \cite{Singh1989,Igarashi1992}, 
but somewhat higher than experimental estimates for \LCO~ \cite{Reehuis2006}.

\textit{Excitation spectra.}
We now turn to spin dynamics, which we calculate within the same MPS 
framework, using the time-dependent variational principle (TDVP)~\cite{Haegeman2011, Haegeman2016}. 
Since we are principally interested in magnon excitations, we concentrate on the transverse structure 
factor~\footnote{Downfolding from the Hubbard model to an effective spin model both the Hamiltonian and the physical observables acquire corrections in powers of $t/U$~\cite{MacDonald1988,Knetter2003} due to charge fluctuations. These operator corrections can modify the expected value of observables such as the staggered magnetization~\cite{Delannoy2005} and vary with $t/U$. For the static or dynamical structure factor $S(\bold{q},\omega)$, an overall renormalization factor $\left(1 - 4\gamma^{(1)}_{\bold{q}} (t/U)^2\right)$ has been reported~\cite{DelannoyThesis2005}. However, given the small effect on the intensities ($\lesssim 5\%$) such a factor has at the $t/U$ values of interest, we have not included it in our calculations.}
\begin{eqnarray}
S_\perp(\bold{q},\omega)
     &=&\int_{-\infty}^{\infty} \frac{dt}{2\pi} \sum_{\bold{r}_i} e^{i(\omega t- \bold{q}\cdot (\bold{r}_i-\bold{r}_j))} \nonumber\\
    &&\langle S^+_i(t) S^-_j (0) + S^-_i(t) S^+_j(0)  \rangle \; .
\label{eq:Sqw.perp}
\end{eqnarray}
In Fig.~\ref{fig:1} we present results obtained for
\mbox{$t/U = 0.102$} [Fig.~\ref{fig:1}(a)], 
\mbox{$t/U =0.141$} [Fig.~\ref{fig:1}(b)], 
and \mbox{$t/U = 0.154$} [Fig.~\ref{fig:1}(c)].   
In all cases, energy is measured in units of $J_1$, and $S_\perp(\bold{q},\omega)$ has been convoluted 
with Gaussian envelope with \mbox{$\sigma_\omega \approx 0.072\ J_1$}, 
such that an infinitely sharp excitation is rendered as Gaussian with FWHM $=0.17\ J_1$.
For comparison, we also show the results of linear spin wave theory (LSWT), 
and a spin wave theory with leading interaction corrections (LSWT+1/$S$) \cite{supplemental.material}, 
calculated for the same parameter set.

The magnon dispersion found in MPS calculations shows clearly defined, linearly-dispersing 
magnons approaching the ordering vector, ${\bf q}_M = (\pi,\pi)$. 
At higher energies, the dispersion is more complicated, and shows a progressive 
evolution as a function of $t/U$.   
Two trends stand out.
The first of these is a reduction in the spin-wave velocity characterizing 
Goldstone modes for ${\bf q} \to {\bf q}_M$.
The second is a softening of excitations at ${\bf q}_\Sigma = (\pi/2,\pi/2)$, 
relative to ${\bf q}_X = (\pi,0)$.
This effect is particularly marked for $t/U = 0.154$ [Fig.~\ref{fig:1}(c)].
For small $t/U$, LSWT+1/$S$ gives a reasonable account of magnon dispersion 
[Fig.~\ref{fig:1}(a)].
However, it fails to capture the softening at $\Sigma$, leading to significant deviations 
from MPS results at larger $t/U$ [Fig.~\ref{fig:1}(c)].


These findings motivate us to introduce an empirical scaling relation for 
ratio of magnon energies, $E_\Sigma/E_X$.
Expressing this in terms of the dominant interactions $J_1$ and $J_c$, 
we find that the results of MPS calculations are well described by 
\begin{equation}
J_c/J_1\approx-E_\Sigma/E_X +1.09 \; ,
\label{eq:emp.relation}
\end{equation}
as illustrated in Fig.~\ref{fig:1}(d).  
This result can easily be reexpressed in terms of $t/U$, 
and used to extract
Hubbard model parameters from experimental measurements of $E_\Sigma$ and $E_X$.
The resulting estimates for CFTD~\cite{Piazza2014}, ${\mathrm{La}}_{2}{\mathrm{CuO}}_{4}$~\cite{Headings2010}, 
$\mathrm{CaCuO}_2$~\cite{Martinelli2022}, 
$\mathrm{Sr}_2\mathrm{IrO}_4$~\cite{Kim2012} and $\mathrm{SrCuO}_2$~\cite{Wang2024} 
are listed in Table~\ref{table:t.U.estimates}.


\begin{table}[t]
\centering
\begin{tabular}[t]{|c|c|c|c|c|}
\hline
& $E_\Sigma$[meV] & $E_X$[meV] & $J_c/J_1$ [Eq.~(\ref{eq:spin.model})] & $t/U$ [Eq.~(\ref{eq:Hubbard.model})] \\
\hline
$\mathrm{CFTD}$~\cite{Piazza2014} & 13.3 & 14.5 & 0.0 & 0.0 \\
${\mathrm{La}}_{2}{\mathrm{CuO}}_{4}$~\cite{Headings2010} & 281 & 323 & 0.22 & 0.102\\
$\mathrm{CaCuO}_2$~\cite{Martinelli2022} & 240 & 375 & 0.45 & 0.141\\
$\mathrm{Sr}_2\mathrm{IrO}_4$~\cite{Kim2012}  & 110 & 205 & 0.55 & 0.154\\
$\mathrm{SrCuO}_2$~\cite{Wang2024} & 191 & 362 & 0.56 & 0.155\\
\hline
\end{tabular}
\caption{
\label{table:t.U.estimates}
Estimates of  model parameters for square-lattice antiferromagnets, obtained using empirical scaling 
relation Eq.~\eqref{eq:emp.relation}.
Corresponding predictions for magnon spectra in ${\mathrm{La}}_{2}{\mathrm{CuO}}_{4}$, $\mathrm{CaCuO}_2$
and $\mathrm{Sr}_2\mathrm{IrO}_4$ are shown in Fig.~\ref{fig:1}(a)-(c).
}
\end{table}


{\it Application to ${\mathrm{La}}_{2}{\mathrm{CuO}}_{4}$}. 
In Fig.~\ref{fig:1}(a) we present a comparison of the magnon spectra found
in MPS calculations for \mbox{$t/U = 0.102$}, and inelastic neutron scattering 
experiments on ${\mathrm{La}}_{2}{\mathrm{CuO}}_{4}$~\cite{Headings2010}.
Results are shown for both the dispersion, characterized by 
$S_\perp(\bold{q}, \omega)$, and 
the intensity of the magnon peak
$
 I(\bold{q}) = S_\perp(\bold{q}, E_{\bf q}), 
$
where $E_{\bf q}$ is the associated magnon energy. 
We find excellent agreement for both quantities across the vast majority of the Brillouin zone.
Nonetheless, for ${\bf q} \approx (\pi,0)$, the intensity measured in experiment shows a small, but 
systematic, reduction relative to simulation.

\begin{figure}[tb]
   \centering
	\includegraphics[width=1\textwidth]{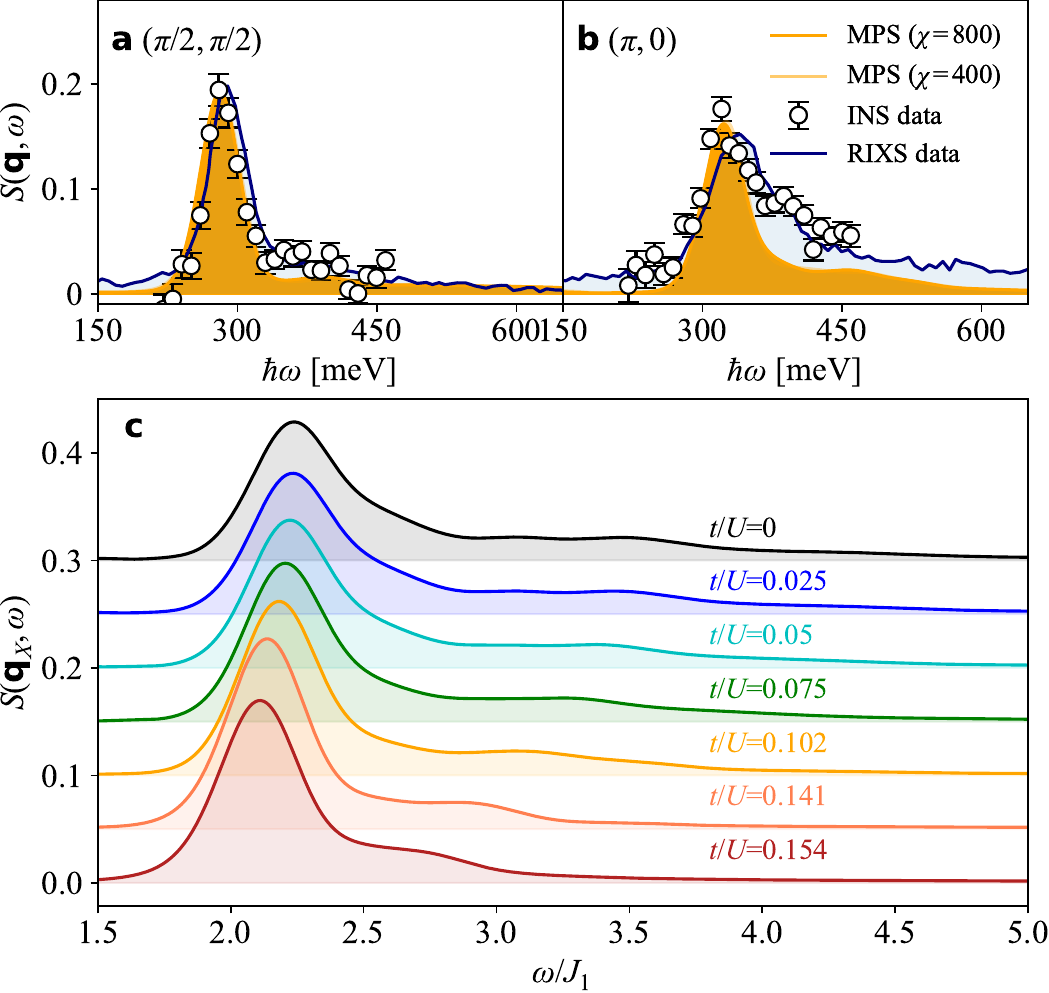}
    \caption{
    Comparison of excitations on the magnetic Brillouin Zone boundary, as found in simulations 
    using matrix product state (MPS), and experiments on \LCO.    
    (a) Dynamical structure factor $S({\bf q},\omega)$ for \mbox{${\bf q}_\Sigma = (\pi/2, \pi/2)$}, 
    showing good agreement between MPS results for \mbox{$t/U = 0.102$},
    inelastic neutron scattering (INS), and resonant inelastic x-ray scattering (RIXS)~\cite{Robarts2021}.
    (b) Equivalent results for \mbox{${\bf q}_X = (\pi, 0)$}, showing agreement on peak position and amplitude, but significant 
    differences at high energy.
    (c) Evolution of $S({\bf q}_X,\omega)$ with increasing $t/U$, showing how spectral weight is redistributed 
    from the high-energy tail found in the Heisenberg model \mbox{[$t/U=0$]}, to lower energies.
    All MPS simulations were carried out for Eq.~(\ref{eq:spin.model}), with maximum bond dimension $\chi = 800$, 
    and energy resolution $0.13\ J_1$, comparable to experiment.
   }
    \label{fig:3}
\end{figure}

Turning to the magnon lineshape, in Fig.~\ref{fig:3} we show a comparison between 
dynamical structure factor measured in INS and RIXS experiments, and calculated using MPS. 
In this case, results are shown for  
$S({\bf q}, \omega)$, defined as
\begin{eqnarray}
S({\bf q}, \omega)
	=  \int_{-\infty}^{\infty} \frac{dt}{2\pi} \sum_{\bold{r}_i} \, e^{i(\omega t- \bold{q}\cdot (\bold{r}_i-\bold{r}_j))} 
     \langle {\bf S}_i(t) \cdot {\bf S}_j (0)  \rangle .
\label{eq:Sqw}
\end{eqnarray}
For \mbox{${\bf q}_\Sigma = (\pi/2,\pi/2)$} [Fig.~\ref{fig:3}(a)], the majority of spectral weight is found 
in the magnon peak, and MPS results provide a good account of experiment.
However, for ${\bf q}_X = (\pi,0)$ [Fig.~\ref{fig:3}(b)], a significant fraction of the spectral weight 
measured in experiment is found in a broad high-energy 
continuum \cite{Headings2010,Braicovich2010,Dean2012,Robarts2021,Betto2021}, 
which has been discussed as potential evidence for spinon excitations \cite{Ho2001, Dean2012, Shao2017}.
While MPS calculations for the effective spin model, 
Eq.~(\ref{eq:spin.model}), correctly 
reproduce the dispersion for ${\bf q} \approx {\bf q}_X$, 
they do not reproduce the lineshape seen in experiment.
What distinguishes them is the continuum at high energies, which manifests as a broad, highly asymmetric 
peak in experiment, and as a weaker, high--energy tail in simulation.  

Calculations of $S({\bf q},\omega)$ for the Heisenberg AF [Eq.~(\ref{eq:spin.model}), $\lim t/U \to 0$], 
find a ``roton minimum'' in the magnon dispersion at ${\bf q}_X = (\pi,0)$ \cite{Syljuasen2000,Sandvik2001,Powalski2015}, 
accompanied by significant spectral weight at high energies \cite{Piazza2014,Shao2017,Powalski2018,Verresen2018}.
We have confirmed that our calculations are consistent with published MPS results for 
the Heisenberg AF \cite{Verresen2018}.
However, with increasing $t/U$, we find that $S({\bf q}_X,\omega)$ exhibits a transfer of spectral weight 
from high to lower energies [Fig.~\ref{fig:3}(c)], leading to the result shown in Fig.~\ref{fig:3}(b).  
These calculations do not, by themselves, resolve whether the high-energy continua 
observed in \LCO, \CCO, and CFTD originates in fractionalized excitations.  
However, since MPS results are robust against changes in bond dimension, cylinder circumference, and 
cylinder geometry \cite{supplemental.material}, we infer that the disagreement between 
experiment and simulation for ${\bf q} \approx {\bf q}_X$  [Fig.~\ref{fig:3}(b)] 
cannot be explained using Eq.~(\ref{eq:spin.model}).
We return to this point below.


{\it Application to $\mathrm{CaCuO}_2$}.  
The infinite-layer cuprate $\mathrm{CaCuO}_2$ is believed to exhibit particularly large cyclic 
exchange \cite{Braicovich2009,Peng2017,Martinelli2022}.
In Fig.~\ref{fig:1}(b) we show magnon spectra found
in MPS calculations for \mbox{$t/U = 0.141$}, as compared with RIXS 
experiments on the $\mathrm{CaCuO}_2$~\cite{Martinelli2022}.
Setting $J_1=172~\text{meV}$, we find reasonable agreement for the dispersion at high energies.   
At low energies, the RIXS spectra show a gap at ${\bf q} = \Gamma$, which has been attributed 
to interlayer coupling~\cite{Martinelli2022}.
Since this is not included in our model, comparison is difficult.
However, the overall distribution of intensity $I(\bold{q})$ shows good 
agreement with experiment, except near \mbox{${\bf q}_X  = (\pi, 0)$}, where experiment 
shows a significant high--energy continuum, similar to that found in ${\mathrm{La}}_{2}{\mathrm{CuO}}_{4}$.   
RIXS data for $\mathrm{CaCuO}_2$ have previously been fitted using LSWT~\cite{Martinelli2022}, 
leading to an estimate $t/U=0.194$, which places $\mathrm{CaCuO}_2$ near the limits of stability 
of $(\pi,\pi)$ N\'eel order [Fig.~\ref{fig:2}(d)].  
As in ${\mathrm{La}}_{2}{\mathrm{CuO}}_{4}$, MPS results suggest that this is a significant overestimate, cf. 
Fig.~\ref{fig:1}(d).   


{\it Application to $\mathrm{Sr}_2\mathrm{IrO}_4$}.  
The quasi-two-dimensional antiferromagnet $\mathrm{Sr}_2\mathrm{IrO}_4$ provides an interesting counterpoint 
to ${\mathrm{La}}_{2}{\mathrm{CuO}}_{4}$, since it has a similar phenomenology \cite{Fujiyama2012,Kim2012},  
but is  known to exhibit strong spin-orbit coupling, with Ir$^{4+}$ moments 
having the character \mbox{$j=1/2$}~\cite{Kim2008,Jackeli2009}.
In Fig.~\ref{fig:1}(c) we show comparison of RIXS data for $\mathrm{Sr}_2\mathrm{IrO}_4$~\cite{Kim2012}
 with MPS calculations for $t/U=0.154$, setting $J_1=96~\text{meV}$~\cite{Fujiyama2012}.
We find excellent agreement for dispersion and intensity of magnon excitations across the entire 
Brillouin zone, consistent with the idea that $\mathrm{Sr}_2\mathrm{IrO}_4$ can be described by a one-band Hubbard 
model \cite{Wang2011}.   
However, it should be noted the RIXS data shown in Fig.~\ref{fig:1}(c) does not have 
sufficiently high resolution to probe any gap at ${\bf q}_M = (\pi,\pi)$ coming from terms breaking
spin-rotation symmetry \cite{Pincini2017}.


{\it Limitations of LSWT}.  
Spin wave theory (SWT) remains an important tool for interpreting magnon spectra  
in experiment~\cite{Coldea2001,Katanin2002,Headings2010,Kim2012,Dean2012,Robarts2021,Martinelli2022}.
Relative to MPS, fits based on low orders of spin wave theory lead to systematic 
overestimates in values of parameters [Fig.~\ref{fig:1}(e)].
This follows from two effects: first that the spin model, Eq.~(\ref{eq:spin.model}), 
is underconstrained, and second that corrections to LSWT from quantum fluctuations 
are of the same scale as those coming from subleading interactions.   


While Eq.~(\ref{eq:spin.model}) has four parameters \mbox{$(J_1, J_2, J_3, J_c)$}, 
magnon spectra for ${\mathrm{La}}_{2}{\mathrm{CuO}}_{4}$ are well-described 
by LSWT calculations for an effective model with only two parameters:  
an AF interaction $J^{\sf eff}_1$ on first-neighbor bonds, 
and a FM  interaction $J^{\sf eff}_2$ on second-neighbor bonds \cite{Toader2005}.
In \cite{supplemental.material} we show how such a two-parameter model can be derived 
from Eq.~(\ref{eq:spin.model}), and parameterized from a one-band Hubbard model, 
within the lowest order of interacting SWT (LSWT + $1/S$).
However, in order to mimic the softening of magnon dispersion at ${\bf q}_\Sigma = (\pi/2,\pi/2)$ 
by quantum fluctuations, $t/U$ must be made artificially large.
And this in turn leads to significant errors in estimates of  
\mbox{$(J_2, J_3, J_c)$} [Fig.~\ref{fig:1}(e)].


In the case of ${\mathrm{La}}_{2}{\mathrm{CuO}}_{4}$, fits based on LSWT overestimate $J_2$, $J_3$, and $J_c$ 
by $\sim 100\%$ [Fig.~\ref{fig:1}(e)].
Including $1/S$ corrections reduces this error to $\gtrsim 40\%$.
Systematic calculations of higher-order corrections 
\cite{Hamer1992,Igarashi1992,Weihong1993,Igarashi2005,Zhitomirsky2013} 
are currently lacking for any realistic model of ${\mathrm{La}}_{2}{\mathrm{CuO}}_{4}$.  
However, recent calculations for Eq.~(\ref{eq:spin.model}) within a 
modified spin wave theory, compare well with MPS results, at the expense of 
a more complicated formalism~\cite{Yamamoto2019}.


{\it What is missing from the spin model?}  
If we take the disagreement between experiment and simulations at face value [Fig.~\ref{fig:3}(b)], 
we are obliged to ask what is missing from our model of magnetism in ${\mathrm{La}}_{2}{\mathrm{CuO}}_{4}$?
One possibility is that the one-band Hubbard model [Eq.~(\ref{eq:Hubbard.model})] remains valid, but 
that charge fluctuations at high order in $t/U$ modify the spin dynamics.    
To rule this out, it would be necessary to calculate directly from the 
Hubbard model, which is beyond the scope of the present work.
Alternatively, the Hubbard model itself might need modification.    
The simplest extension would be 
hopping $t^\prime$ on second-neighbor 
bonds~\cite{Delannoy2009}, a term which has been argued to  play an important role 
in superconductivity \cite{Qin2020, Xu2024}.
Further refinements include spin-orbit coupling~\cite{Thio1988,Coffey1990,Shekhtman1992,Mazurenko2005},
generalization to a three-band model \cite{Zhang1988,Kowalski2021,Aligia2018},
or coupling to phonons. 
To distinguish between these alternatives, it may also be necessary to revisit experiment.
We leave these as questions for future work.

\textit{Summary and conclusions.}
%
In this Letter, we have used calculations based on matrix product states (MPS) 
to characterize the magnon spectra of cuprate materials, starting from the model 
commonly used to fit experiments: an effective spin model with four-spin exchange 
[Eq.~(\ref{eq:spin.model})], which descends from a one-band Hubbard model 
[Eq.~(\ref{eq:Hubbard.model})].
We have made explicit comparison with experimental results for 
${\mathrm{La}}_{2}{\mathrm{CuO}}_{4}$ [Fig.~\ref{fig:1}(a)], 
$\mathrm{CaCuO}_2$ [Fig.~\ref{fig:1}(b)], and 
$\mathrm{Sr}_2\mathrm{IrO}_4$ [Fig.~\ref{fig:1}(c)], 
finding generally good agreement with the measured magnon dispersion 
throughout the Brillouin zone.
We also find good agreement for magnon lineshape, 
except near ${\bf q}_X = (\pi,0)$, where experiment 
reveals a high-energy continuum which is not well described 
by simulations [Fig.~\ref{fig:3}(b)]. 
We estimate the ratio $t/U$ that characterizes a one-band Hubbard in each 
case [Table~\ref{table:t.U.estimates}], finding values which are systematically 
lower than those obtained in published fits to linear spin wave theory (LSWT).  
We also introduce a method of estimating $t/U$ directly from measured 
magnon energies at ${\bf q}_X = (\pi,0)$ and ${\bf q}_\Sigma = (\pi/2,\pi/2)$ 
[Fig.~\ref{fig:1}(d)].
%


These results suggest two main conclusions.
First, that fitting the magnon dispersion using linear spin wave theory 
leads to systematic errors in estimates of in values of exchange parameters, 
and corresponding overestimates of $t/U$.   
And second that, suitably parameterized, the ``standard'' model 
describes some, but not all of the properties of the magnon 
spectrum in cuprate antiferrimagnets, providing a good
overall account of dispersion, but failing to capture the 
continuum observed at high energies.

\textit{Acknowledgments.}
 We thank Stephen Hayden for discussions and sharing experimental data. 
 We thank Radu Coldea, Bruce Gaulin, Michel Gingras, Hui Shao, Tokuro Shimokawa, Ruben Verresen 
 and Ling Wang for helpful discussions and comments. 
 This research is supported by the Theory of Quantum Matter Unit at Okinawa Institute of Science and Technology (OIST), the Japan Society for the Promotion of Science (JSPS) Research Fellowships for Young Scientists (Grant No. 23KJ2136), and JSPS KAKENHI (Grant No. JP22K14008).
  Work at the University of Windsor (JGR) was funded by the Natural Sciences and Engineering Research Council of Canada (NSERC) (Funding Reference No. RGPIN-2020-04970).
 We acknowledge the use of computational resources of the Scientific Computing 
 section of the Research Support Division at OIST
and of the ISSP Supercomputer Center at the University of Tokyo.


\bibliography{main.bib}



\section*{Supplementary material}

# Supplemental material: Magnon Spectra of Cuprates beyond Spin Wave Theory

Jiahui Bao,[1, *] Matthias Gohlke,[1] Jeffrey G. Rau,[2] and Nic Shannon[1]

[1]*Theory of Quantum Matter Unit, Okinawa Institute of Science and Technology Graduate University, Onna-son, Okinawa 904-0495, Japan*
[2]*Department of Physics, University of Windsor, 401 Sunset Avenue, Windsor, Ontario, N9B 3P4, Canada*
(Dated: July 29, 2024)

## I. FROM HUBBARD MODEL TO SPIN MODEL

Our starting point for understanding the magnetism of $La_2CuO_4$, and the other materials consider in the main text, is a set of localized electrons on a square lattice. The general problem of exchange of localized Fermions with non-relativistic (i.e. spin–rotationally invariant) interactions was first considered by Dirac as a general paradigm for quantum many body physics [1], and later revisited by Thouless in the context of $^3$He [2]. The Hamiltonian describing these localized fermions can be written as a sum of permutation operators

$$\mathcal{H}_{\mathsf{P}} = -\sum_\lambda (-1)^{p_\lambda} \tilde{J}^\lambda \mathcal{P}^\sigma_\lambda \,, \tag{S1}$$

where $\lambda$ counts all possible cyclic permutations of spins $\mathcal{P}^\sigma_\lambda$, $\tilde{J}^\lambda$ is corresponding matrix element (exchange integral), and the parity $p_\lambda = \pm 1$ depends on whether the number of fermions permuted is odd or even. It follows from Fermi statistics [1], that permutations of an even number of spins lead to antiferromagnetic interactions, while permutation of an odd number of spins lead to ferromagnetic interactions. In this work, we consider permutations of up to 4 spins, leading to an effective Hamiltonian [3, 4]

$$\mathcal{H}_{\mathsf{P}} = \tilde{J}^{(2)}_1 \sum_{\langle ij \rangle_1} \mathcal{P}_{ij} + \tilde{J}^{(2)}_2 \sum_{\langle ij \rangle_2} \mathcal{P}_{ij} + \tilde{J}^{(2)}_3 \sum_{\langle ij \rangle_3} \mathcal{P}_{ij} - \tilde{J}^{(3)} \sum_{\langle ijk \rangle} [\mathcal{P}_{ijk} + \mathcal{P}^{-1}_{ijk}] + \tilde{J}^{(4)} \sum_{\langle ijkl \rangle} [\mathcal{P}_{ijkl} + (\mathcal{P}_{ijkl})^{-1}], \tag{S2}$$

where $\tilde{J}^{(2)}_n$ characterizes exchange of 2 spins on an $n^{\text{th}}$-neighbour bond, and $\tilde{J}^{(3)}$ and $\tilde{J}^{(4)}$ are three- and four-spin cyclic exchanges, defined on the cycles shown in Fig. S1.

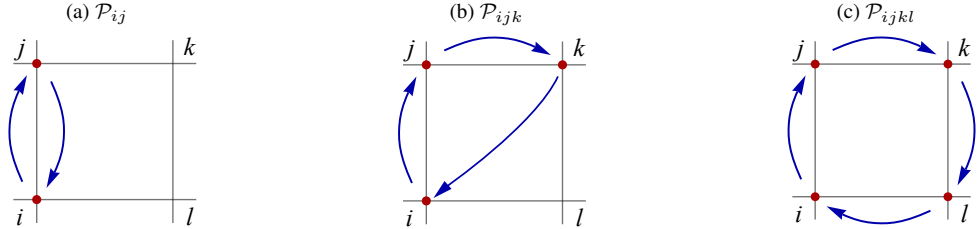

Figure S1. Cyclic permutations of spins on the square lattice, as found in $\mathcal{H}_{\mathsf{P}}$ [Eq. S2]. (a) Cyclic permutation of two spins, as represented by the operator $\mathcal{P}_{ij}$. (b) Cyclic permutation of three spins, as represented by the operator $\mathcal{P}_{ijk}$. (c) Cyclic permutation of four spins, as represented by the operator $\mathcal{P}_{ijkl}$. The inverse of each operator $\mathcal{P}^{-1}_\lambda$ corresponds to the cycle taken with the opposite sense.

The exchange integrals $\tilde{J}^\lambda$ which parameterize $\mathcal{H}_{\mathsf{P}}$ can be determined in many different ways, and will in general depend on the level of sophistication of the treatment of the underlying Mott insulator. In this work, we consider the simplest possible model of $La_2CuO_4$ namely, the one-band Hubbard model

$$\mathcal{H}_{\mathsf{U}} = -t \sum_{\langle ij \rangle_1 \sigma} \left[ c^\dagger_{i\sigma} c_{j\sigma} + \text{h.c.} \right] + U \sum_i c^\dagger_{i\uparrow} c_{i\uparrow} c^\dagger_{i\downarrow} c_{i\downarrow} \,, \tag{S3}$$

at half filling, on a square lattice, for $t/U \ll 1$. An effective spin model can be derived from $\mathcal{H}_{\mathsf{U}}$ through a perturbative expansion in $t/U$, either by working with the operators for electrons [5–7], or by enumerating the processes which contribute

---

* jiahui.bao@oist.jp

to each cyclic permutation $\tilde{J}^\lambda$ in the spin basis [3, 4]. Either approach leads to the same model. Here we continue the logic cyclic–permutations, in which case the parameters of $\mathcal{H}_\mathrm{P}$ [Eq. (S2)] are given by,

$$\tilde{J}_1^{(2)} = 2(t^2/U)(1 + 4t^2/U) \, , \tag{S4a}$$

$$\tilde{J}_2^{(2)} = 12t^4/U^3 \, , \tag{S4b}$$

$$\tilde{J}_3^{(2)} = 2t^4/U^3 \, , \tag{S4c}$$

$$\tilde{J}^{(3)} = 10t^4/U^3 \, , \tag{S4d}$$

$$\tilde{J}^{(4)} = 20t^4/U^3 \, . \tag{S4e}$$

We note that it is also possible to parameterize $\mathcal{H}_\mathrm{P}$ starting from a 3–band Hubbard model [8]. However the resulting expressions for $\tilde{J}^\lambda$ depend on many more parameters, and in this work we will restrict ourself to the one–band case.

For purposes of some calculations, e.g. exact diagonalization, it makes sense to work directly with a Hamiltonian expressed in terms of permutation operators. However in other cases, e.g. spin wave theory, it is more convenient to work with a model defined in terms of spins. In the absence of magnetic field [1, 2], cyclic permutations can be decomposed in terms of permutations of pairs of spins

$$\mathcal{P}_{ijkl} = \mathcal{P}_{ij}\mathcal{P}_{jk}\mathcal{P}_{kl} \, , \tag{S5a}$$

$$\mathcal{P}_{ijk} = \mathcal{P}_{ij}\mathcal{P}_{jk} \, , \tag{S5b}$$

Substituting this result in Eq. (S2), and using the relation

$$\mathcal{P}_{ij} = (1 + 4\mathbf{S}_i \cdot \mathbf{S}_j)/2 \, , \tag{S6}$$

we arrive at the spin model

$$\mathcal{H}_\sigma = (2\tilde{J}_1^{(2)} - 8\tilde{J}^{(3)} + 2\tilde{J}^{(4)}) \sum_{\langle ij \rangle_1} \mathbf{S}_i \cdot \mathbf{S}_j + (2\tilde{J}_2^{(2)} - 4\tilde{J}^{(3)} + \tilde{J}^{(4)}) \sum_{\langle ij \rangle_2} \mathbf{S}_i \cdot \mathbf{S}_j + 2\tilde{J}_3^{(2)} \sum_{\langle ij \rangle_3} \mathbf{S}_i \cdot \mathbf{S}_j$$
$$+ 4\tilde{J}^{(4)} \sum_{\langle ijkl \rangle} (\mathbf{S}_i \cdot \mathbf{S}_j)(\mathbf{S}_k \cdot \mathbf{S}_l) + (\mathbf{S}_j \cdot \mathbf{S}_k)(\mathbf{S}_l \cdot \mathbf{S}_i) - (\mathbf{S}_i \cdot \mathbf{S}_k)(\mathbf{S}_j \cdot \mathbf{S}_l). \tag{S7}$$

Finally, we collect parameters as

$$J_1 = 2\tilde{J}_1^{(2)} - 8\tilde{J}^{(3)} + 2\tilde{J}^{(4)} = 4\frac{t^2}{U}\left(1 - 6\frac{t^2}{U^2}\right) \, , \tag{S8a}$$

$$J_2 = 2\tilde{J}_2^{(2)} - 4\tilde{J}^{(3)} + \tilde{J}^{(4)} = 4\,\frac{t^4}{U^3} \, , \tag{S8b}$$

$$J_3 = 2\tilde{J}_3^{(2)} = 4\,\frac{t^4}{U^3} \, , \tag{S8c}$$

$$J_c = 4\tilde{J}^{(4)} = 80\,\frac{t^4}{U^3} \, , \tag{S8d}$$

to obtain the spin model $\mathcal{H}_\sigma$ [Eq. (1) of the main text], which is used as the starting point for all the calculations in this work.

We note that, as defined above, the 4-spin interaction $J_c$ has a different dimension from the other three exchange interactions, scaling as $[J_c S^2] \sim [J_1]$. This problem can be resolved by interpreting the factor 4 within the definition of $J_c$ as $1/S^2$, and reparameterizing the model in terms of

$$K = \tilde{J}^{(4)} = J_c S^2 \, . \tag{S9}$$

We return to this point below in the context of spin wave theory, where it is required to achieve an homogenous expansion of fluctuations.

## II. DETAILS OF NUMERICAL SIMULATIONS

### A. MPS calculations

The numerical simulations of the ground state and the dynamical structure factor are carried out utilizing the matrix product state (MPS) framework [9], and the TeNPy library [10]. In order to use the one-dimensional tensor train of the MPS for two-dimensional systems, we wrap the square lattice on a cylinder and wind around the MPS structure. We use density matrix

renormalization group (DMRG) [11] assuming an infinite MPS ansatz [12] to obtain the ground state of the effective spin model [Eq. (S17)] and its properties such as energy, staggered magnetization and equal-time correlations. Ground state calculations are done using a bond dimension of up to $\chi = 1600$ and taking advantage of $S^z$-conservation.

Having characterized the ground state, we compute the spin dynamics using the time-dependent variational principle (TDVP) [13, 14] on finite cylinders. Except where specified, calculations were carried out using a single–site implementation of TDVP, with a time step $\Delta t = 0.1\,J_1^{-1}$. In order to keep the computational costs within reasonable bounds, we obtain the ground state on the finite cylinder with $\chi = 800$ and keep $\chi$ constant throughout the time evolution. The maximum time is set to $t_{\max} = 40\,J_1^{-1}$ for higher resolution plots [Fig. 1(a-c) in the main text], while simulations with $t_{\max} = 10\,J_1^{-1}$ are sufficient to extract the position of the single magnon excitation.

To eliminate the Gibbs phenomenon caused by the finite duration of the time evolution, we use linear prediction to extrapolate the time series to long times [15–17], followed by multiplying with a Gaussian envelope $e^{-t^2/2\sigma_t^2}$. After Fourier transformation, the effective energy resolution is then given by

$$\sigma_\omega = 1/\sigma_t = \lambda_c/t_{\max} \tag{S10}$$

where we choose $\lambda_c = 3 \sim 4$ for all computations.

Further technical details of the simulations, as well as checks on convergence, are documented below.

## B. Cylinder Geometry

The finite circumference of the cylinder results in discrete momenta $q_y = 2\pi/L_y$ in reciprocal space. In order to access to different momentum cuts, we use two distinct cylinder geometries. In particular, the non-rotated geometry with the edges of the square lattice being parallel to the cylinder axis [Fig. S2a], and the geometry subject to a $\pi/4$ rotation relative to the cylinder axis [Fig. S2b]. The rotated square lattice has a two–site unit cell, resulting in a zigzag path of the circumference of the cylinder. Throughout the text, we define the circumference $L_y$ in Manhattan distance, which is the minimal number of jumps along nearest neighbor bonds required to go around the cylinder.

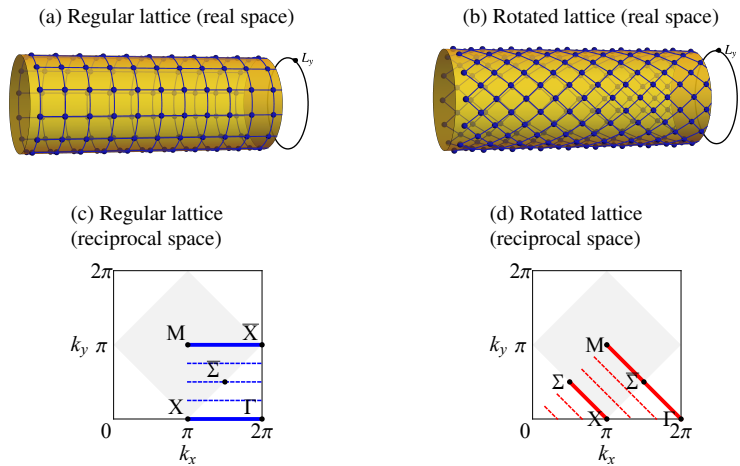

Figure S2. Cylinder geometries and its square lattice wrapping utilized in matrix product state (MPS) calculations. (a) Regular geometry with all bonds either parallel or perpendicular to the cylinder axis. (b) Rotated geometry with bonds at angle $\pi/4$ to the cylinder axis. (c) Cuts through reciprocal space accessible through a regular cylinder geometry with $L_y = 8$. The grey shaded area denotes the magnetic Brillouin zone associated with $(\pi, \pi)$ Néel order. (d) Equivalent cuts through reciprocal space for the rotated cylinder geometry with $L_y = 12$.

In Fig. S2c and Fig. S2d we illustrate the Brillouin zone and the accessible momenta cuts. Both geometries are appropriate for hosting the Néel ordered ground state. In order to access some of the experimentally relevant paths in reciprocal space, we make use of $\mathbf{k}_{\bar{X}} = (0, \pi)$ and $\mathbf{k}_{\bar{\Sigma}} = (-\pi/2, \pi/2)$ instead of $\mathbf{k}_X = (\pi, 0)$ and $\mathbf{k}_\Sigma = (\pi/2, \pi/2)$. We have confirmed that finite–size effects are small at large $L_y$ and discrepancies due to the change of coordinates are negligible.

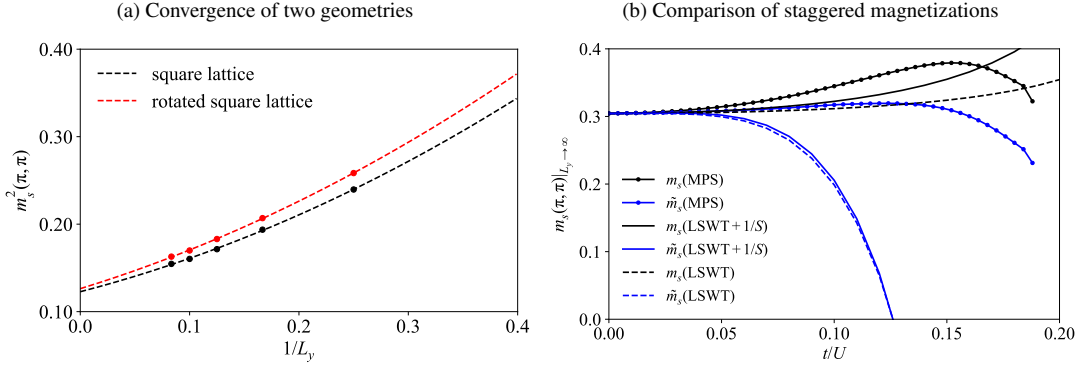

Figure S3. Staggered magnetization (squared) as a function of cylinder circumference $L_y$ and its extrapolation to $L_y \to \infty$ (a) Extrapolation of $m_s(\pi, \pi)$ [Eq. (S11)], for $t/U = 0.1$, following Eq. (S12), showing convergence for cylinders with regular and rotated geometries [Fig. S2]. (b) Staggered magnetization in the thermodynamic limit as a function of $t/U$, obtained from extrapolating the results of the regular square geometry. We include the staggered magnetization extracted from MPS simulations $m_s$ (black symbols), $\bar{m}_s$ [Eq. (S14)] including a correction for the effect of charge fluctuations (blue symbols), and estimates from linear (LSWT) and interacting (LSWT + 1/S) spin–wave calculations (solid and dashed lines, respectively), cf. Sec. III.

## C. Calculation of staggered magnetization

The staggered magnetization

$$m_s^2(\pi, \pi) = \frac{1}{N^2} \sum_{\langle ij \rangle} e^{i(\pi,\pi) \cdot (\mathbf{r}_i - \mathbf{r}_j)} \langle \mathbf{S}_i \cdot \mathbf{S}_j \rangle \,, \tag{S11}$$

is calculated via the equal-time two-point correlators $\langle \mathbf{S}_i \cdot \mathbf{S}_j \rangle$ on a $N = L_y \times L_y$ rectangular patch in the center of an infinite cylinder.

Within a Néel state, the finite–size corrections to $m_s^2(\pi, \pi)$ are controlled by fluctuations of linearly–dispersing magnons at long wavelength, and are polynomial in $1/L$

$$m_s^2(\mathbf{q})|_L = m_s^2(\mathbf{q})|_\infty + a/L + b/L^2 + \mathcal{O}(1/L^3) \,. \tag{S12}$$

(cf. e.g. [18]). In Fig. S8a, we show that the staggered magnetizations found for cylinders with regular and rotated lattice converge onto very similar values for $L \to \infty$ when interpolated in this way. In this case, results are obtained for $t/U = 0.1$.

In Fig. S8b, we show how estimates of the staggered magnetization evolve as a function of $t/U$. In the limit of a nearest–neighbour Heisenberg model, $t/U \to 0$, we find

$$m_s(\pi, \pi) \approx \sqrt{0.093} \approx 0.304 \tag{S13}$$

consistent with published estimates [19–21].

We estimate the corrections to the staggered magnetization which arise from the leading order of charge fluctuations [22],

$$\bar{m}_s = (1 - 2zt^2/U^2)m_s \,, \tag{S14}$$

where $z = 4$ is coordination number of the square lattice. As illustrated in Fig. S8b, charge fluctuations significantly renormalise the staggered magnetization for larger values of $t/U$,

For comparison, we include the (renormalized) staggered magnetization found in linear spin wave theory (LSWT) and spin wave theory with leading $1/S$ corrections (LSWT + 1/S). Once quantum effects are taken into account, the renormalized staggered magnetization remains a convex function of $t/U$. This contrasts with the conclusion reached on the basis of spin wave theory alone in Ref. 22.

## D. Convergence of calculations of dynamics

The fidelity of excitation spectra obtained by the TDVP algorithm in evolving an MPS depends on several factors. Initially, we check the convergence of the TDVP setup utilized. For the accuracy of TDVP algorithms, one/two-site and time step are

two predominant factors. In this work, we use a single-site TDVP accompanied by a time step $\Delta t = 0.1\ J_1^{-1}$. In Fig. S4 we show a comparison of the real-time correlation functions found employing a two-site TDVP or $\Delta t = 0.05\ J_1^{-1}$. For a $100 \times 8$ square-lattice cylinder, with $\chi = 800$ and $t/U = 0.1$, the results exhibit coherence up to $t_{max} = 30\ J_1^{-1}$, roughly $3\times$ that needed to obtain reasonable accuracy in calculations of magnon dispersion.

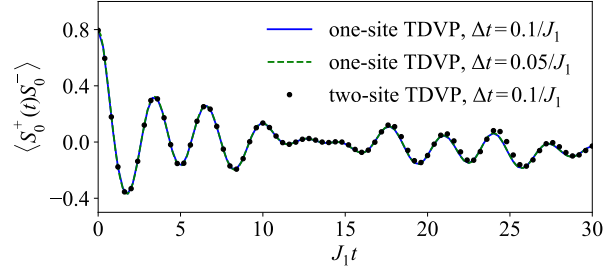

Figure S4. Check on reliability of calculations of dynamics based on time–dependent variational principle (TDVP). The spin autocorrelation function $\langle S_0^+(t)S_0^-(0)\rangle$ found in dynamical MPS simulations is plotted for one– and two–site TDVP ansätze, and for different values of time step $\Delta t$. Good agreement is found up to time $t = 30J_1^{-1}$, establishing the reliability of the results shown in the main text. Simulations were performed on a finite cylinder of dimension $100 \times 8$ spins with a regular square lattice, with a maximum bond dimension $\chi = 800$, for model parameter is $t/U = 0.1$.

The results shown in the main text revolve around MPS calculations of the dynamical structure factor

$$S(\mathbf{q}, \omega) = \int_{-\infty}^{\infty} \frac{dt}{2\pi} \sum_{\mathbf{r}_i} e^{i(\omega t - \mathbf{q} \cdot (\mathbf{r}_i - \mathbf{r}_j))} \langle \mathbf{S}_i(t) \cdot \mathbf{S}_j(0) \rangle. \tag{S15}$$

and of the the transverse structure factor

$$S_\perp(\mathbf{q}, \omega) = \int_{-\infty}^{\infty} \frac{dt}{2\pi} \sum_{\mathbf{r}_i} e^{i(\omega t - \mathbf{q} \cdot (\mathbf{r}_i - \mathbf{r}_j))} \langle S_i^+(t)S_j^-(0) + S_i^-(t)S_j^+(0) \rangle . \tag{S16}$$

which was used to extract single–magnon dispersion. The results obtained can in principle depend on a number of factors beyond the setup of TDVP, including the cylinder geometry and circumference, maximum bond dimension, and maximum evolution time $t_{max}$.

As a first check on the reliability of our results, we confirm that estimates of the single magnon dispersion are not strongly affected by cylinder geometry or circumference. We concentrate on the ratio of magnon energies at $\mathbf{q}_X = (\pi, 0)$ and $\mathbf{q}_\Sigma = (\pi/2, \pi/2)$, since this provides the basis of the empirical scaling relation described in the main text. Direct comparison can only be made for cylinders commensurate with these wave vectors. In practice, this restricts the check to regular cylinders with $L_y = 4, 8$ and cylinders with a rotated geometry with $L_y = 8, 12$.

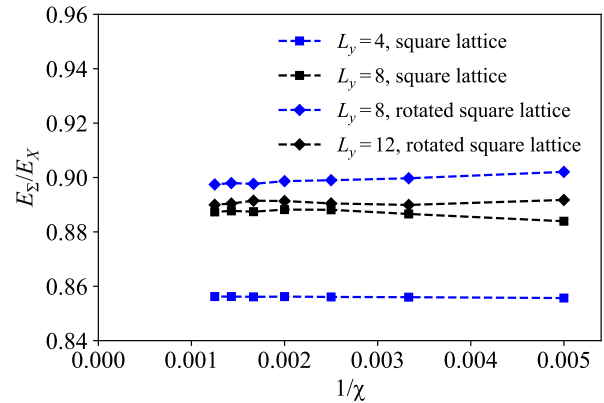

Figure S5. Ratio of magnon energies $E_\Sigma/E_X$ as a function of the inverse bond–dimension $1/\chi$ for cylinders of different circumferences $L_y$, at $t/U = 0.1$, and with $t_{max} = 10\ J_1^{-1}$. Results are shown for regular and rotated square lattices with cylinder circumferences $L_y = 4, 8$ and $L_y = 8, 12$, respectively. Good agreement is found between the largest cylinder of each type, $L_y = 8$ (black squares), and $L_y = 12$ (black diamonds), with little dependence on bond dimension.

In Fig. S5, we show the energy ratios $E_\Sigma/E_X$ obtained from simulations of $S_\perp(\mathbf{q}, \omega)$ [Eq. (S16)] at $\mathbf{q}_\Sigma$ and $\mathbf{q}_X$, with maximum bond dimension in the range $\chi \in [200, 800]$, for $J_c/J_1 = 0.2$ [$t/U = 0.102$], with $t_{max} = 10\ J_1^{-1}$. Generally, there exists only little dependence on $\chi$ and very good agreement between $E_\Sigma/E_X$ on the regular lattice with $L_y = 8$ (as used in the main text) and on the rotated lattice with $L_y = 12$. Hence, we infer that for these larger cylinders, finite-size effects do not significantly affect estimates of magnon energies.

We now explore whether a simulation time of $t_{max} = 10\ J_1^{-1}$, as used in the main text, is sufficient to accurately determine excitation spectra. In Fig. S6 we show results for the transverse structure factor $S_\perp(\mathbf{q}, \omega)$ [Eq. (S16)], calculated for $t/U = 0.1$, at wavevectors $\mathbf{q}_\Sigma$ and $\mathbf{q}_X$. The bond dimension is kept constant at $\chi = 800$, while $t_{max}$ ranges from $10\ J_1^{-1}$ to $40\ J_1^{-1}$. While results at longer simulation time (and therefore higher resolution in energy) reveal new structure at high energies, the energy of the single magnon peaks at both wavevectors prove insensitive to $t_{max}$. We conclude that simulations with $t_{max} = 10\ J_1^{-1}$ provide an accurate estimate of magnon dispersion, while reducing the computational cost significantly.

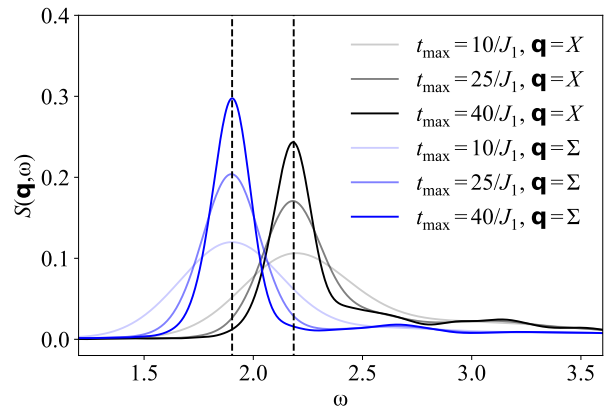

Figure S6. Dependence of single magnon energy estimates on the maximum duration $t_{max}$ of the time evolution. Results are shown for the transverse structure factor $S_\perp(\mathbf{q}, \omega)$ [Eq. (S16)], at $\mathbf{q}_X = (\pi, 0)$ and $\mathbf{q}_\Sigma = (\pi/2, \pi/2)$, calculated using a time series with $t_{max} = 10, 25, 40\ J_1^{-1}$. While the details of the lineshape show some dependence on $t_{max}$, coming from the resolution of simulations, estimates of the single magnon energy remain constant. All calculations are carried out for the spin model Eq. (S17) with $t/U = 0.1$ and $\chi = 800$.

### E. Dependence of predictions for dynamical structure factor oncylinder geometry

The dynamical structure factors shown in the main text are calculated using the regular lattice geometry [Fig. S2a] with $L_y = 8$. As illustrated in Fig. 3 of the main text, these show little change between results obtained for a maximum bond dimension of $\chi = 400$ and those for $\chi = 800$, the default value for calculations of dynamics. However, we provide an additional check on how cylinder circumference and geometry affect the lineshape obtained in simulation.

In Fig. S7 we show results for $S(\mathbf{q}, \omega)$ [Eq. (S15)], for $\mathbf{q} = \mathbf{q}_X$, obtained for cylinders with regular lattice geometry and a circumference of $L_y = 8$, and rotated lattice geometry [Fig. S2b], with a circumference $L_y = 12$. Simulations were carried out for parameters relevant to $\mathrm{La_2CuO_4}$, i.e. $t/U = 0.102$. And order to gain access to possible fine structure coming from finite–size effects, time evolution was continued up to $t_{max} = 40\ J_{-1}$, to give an energy resolution $0.072\ J_1$. The results obtained for the two cylinders are nearly identical, showing small differences in the distribution of spectral weight, which are obscured by the energy–resolution used of the calculations shown in the main text. On these grounds, we conclude that the results shown in the main text are robust against effects of finite size and/or cylinder geometry.

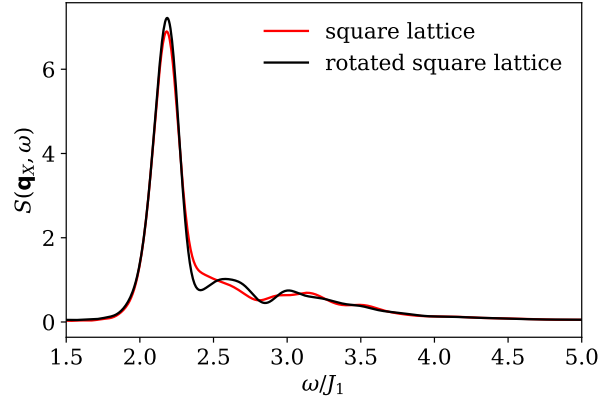

Figure S7. Cylinder geometry dependence of the dynamical structure factor. Results are shown for $S(\mathbf{q}_X, \omega)$ [Eq. (S15)], calculated for cylinders with both regular lattice geometry [Fig. S2a, $L_y = 8$], and rotated lattice geometry [Fig. S2b, $L_y = 12$], for $t_{\max} = 40\ J^{-1}$, giving an energy resolution $0.072\ J_1$. The magnon lines and high energy "continuum" found for the two different cylinder geometries agree up to small differences in the overall distribution of spectral weight. All calculations are carried out for the spin model Eq. (S17), for parameters relevant to $La_2CuO_4$ [$t/U = 0.102$], and maximum bond dimension $\chi = 800$.

## III. SPIN WAVE THEORY AND DERIVATION OF EFFECTIVE 2–PARAMETER MODEL

In what follows we provide a concise summary of spin wave calculations, including leading $1/S$ corrections, as described in the main text. These results parallel earlier discussions in the literature [23], with some additional comments on the form of $1/S$ corrections, and their interpretation in terms of a linear theory for an effective 2–parameter model. Results are presented in such a way as to be easily adapted to the analysis of experiment.

### A. Quantization of magnons and linear spin wave theory (LSWT)

Our starting point for calculations is the effective spin model used in the main text

$$\mathcal{H}_\sigma = J_1 \sum_{\langle ij \rangle_1} \mathbf{S}_i \cdot \mathbf{S}_j + J_2 \sum_{\langle ij \rangle_2} \mathbf{S}_i \cdot \mathbf{S}_j + J_3 \sum_{\langle ij \rangle_3} \mathbf{S}_i \cdot \mathbf{S}_j$$
$$+ \frac{K}{S^2} \sum_{\langle ijlk \rangle} \left[ (\mathbf{S}_i \cdot \mathbf{S}_j)(\mathbf{S}_k \cdot \mathbf{S}_l) + (\mathbf{S}_j \cdot \mathbf{S}_k)(\mathbf{S}_l \cdot \mathbf{S}_i) - (\mathbf{S}_i \cdot \mathbf{S}_k)(\mathbf{S}_j \cdot \mathbf{S}_l) \right] . \tag{S17}$$

where, in order to obtain a homogenous expansion in powers of $S$, we have introduced the parameter

$$K = J_c S^2 = J_c/4 \quad [S = 1/2] . \tag{S18}$$

The validity of this regularization can easily be confirmed by performing a spin wave expansion about the ground state found for exclusively FM interactions, where the one-magnon excitations can be determined exactly. Values of the parameters $J_1$, $J_2$, $J_3$ and $K$, are taken from a one–band Hubbard model through Eq. (S8).

We assume collinear Néel order with ordering vector $(\pi, \pi)$, and spins aligned along the x–axis. To simplify the analysis, we introduce new spin operators rotated by $\pi$ about $\hat{\mathbf{z}}$ on one sublattice, such that

$$\mathbf{S}_i = (-1)^i (\mathsf{S}_i^z \hat{\mathbf{x}} + \mathsf{S}_i^x \hat{\mathbf{y}}) + \mathsf{S}_i^y \hat{\mathbf{z}} . \tag{S19}$$

Spin excitations in the rotated basis can then be quantized through a Holstein–Primakoff transformation

$$\mathsf{S}_i^+ = \sqrt{2S} \left[ 1 - \frac{n_i}{2S} \right]^{1/2} a_i \approx \sqrt{2S} \left[ 1 - \frac{n_i}{4S} + \dots \right] a_i , \tag{S20a}$$

$$\mathsf{S}_i^- = \sqrt{2S} a_i^\dagger \left[ 1 - \frac{n_i}{2S} \right]^{1/2} \approx \sqrt{2S} a_i^\dagger \left[ 1 - \frac{n_i}{4S} + \dots \right] , \tag{S20b}$$

$$\mathsf{S}_i^z = S - n_i , \tag{S20c}$$

in terms of Bosons

$$\left[a_i, a_j^\dagger\right] = \delta_{ij} \ , \tag{S21}$$

with number density

$$n_i = a_i^\dagger a_i \ , \tag{S22}$$

where, as in the main text, we set $\hbar \equiv 1$. From this starting point, noting that the Boson operator $a$ has dimension $\sim \sqrt{S}$, we can define a operator expansion of the different terms arising in the Hamiltonian in powers of the dimensionless parameter $n/4S$. This is conventionally called a "$1/S$" expansion.

The expansion of the exchange term $\mathbf{S}_i \cdot \mathbf{S}_j$ depends on which sublattice the sites $i$ and $j$ belong to. For a general bond this can be written

$$\mathbf{S}_i \cdot \mathbf{S}_j \approx \begin{cases} -S^2 + Sf_{ij} + F_{ij}, & (-1)^i \neq (-1)^j \ , \\ +S^2 + Sg_{ij} + G_{ij}, & (-1)^i = (-1)^j \ , \end{cases} \tag{S23}$$

where we retain terms to $\mathcal{O}(n^2/S^2)$, and introduce the operators

$$f_{ij} \equiv +(n_i + n_j) - (a_i a_j + a_i^\dagger a_j^\dagger) \ , \tag{S24a}$$

$$g_{ij} \equiv -(n_i + n_j) + (a_i a_j^\dagger + a_i^\dagger a_j) \ , \tag{S24b}$$

$$F_{ij} \equiv -n_i n_j + \frac{1}{4}(n_i a_i a_j + a_i n_j a_j + a_i^\dagger n_i a_j^\dagger + a_i^\dagger a_j^\dagger n_j) \ , \tag{S24c}$$

$$G_{ij} \equiv +n_i n_j - \frac{1}{4}(a_i a_j^\dagger n_j + n_i a_j a_j^\dagger + a_i^\dagger n_j a_j + a_i^\dagger n_j a_j) \ . \tag{S24d}$$

If we keep only the terms $f_{ij}$ and $g_{ij}$ in Eq. (S23), the resulting theory is quadratic in Bose operators, and reduces to the LSWT

$$\mathcal{H}_{\mathsf{LSWT}} = E_0 + S\Big[(J_1 - 2K)\sum_{\langle ij\rangle_1} f_{ij} + (J_2 - K)\sum_{\langle ij\rangle_2} g_{ij} + J_3\sum_{\langle ij\rangle_3} g_{ij}\Big] + \mathcal{O}\left(\frac{E_0}{S^2}\right) \ , \tag{S25}$$

where

$$E_0 = NS^2[-2J_1 + 2J_2 + 2J_3 + K] \tag{S26}$$

is the classical ground state energy. The model can now be solved by Fourier transforming the magnon operators

$$a_{\mathbf{k}} = \frac{1}{\sqrt{N}}\sum_i e^{-i\mathbf{k}\cdot\mathbf{r_i}}a_i \ , \tag{S27}$$

to give

$$\mathcal{H}_{\mathsf{LSWT}} = E_0 + \sum_{\mathbf{k}} \left(a_{\mathbf{k}}^\dagger \ \ a_{-\mathbf{k}}\right) \begin{pmatrix} A_0(\mathbf{k}) & B_0(\mathbf{k}) \\ B_0(\mathbf{k}) & A_0(\mathbf{k}) \end{pmatrix} \begin{pmatrix} a_{\mathbf{k}} \\ a_{-\mathbf{k}}^\dagger \end{pmatrix} + \dots \ , \tag{S28}$$

where

$$A_0(\mathbf{k}) = 4S(J_1 - 2K) + (J_2 - K)[\gamma_2(\mathbf{k}) - 1] + J_3[\gamma_3(\mathbf{k}) - 1] \ , \tag{S29a}$$

$$B_0(\mathbf{k}) = -4S(J_1 - 2K)\gamma_1(\mathbf{k}) \ , \tag{S29b}$$

and the form factor associated with $n^{\mathrm{th}}$–neighbour bonds is given by

$$\gamma_n(\mathbf{q}) = \frac{1}{2}\sum_{\{\boldsymbol{\delta}_n\}} \cos(\mathbf{q}\cdot\boldsymbol{\delta}_n) \ , \tag{S30}$$

where $\{\boldsymbol{\delta}_n\}$ is the associated set of lattice vectors.

Written in this form, the Hamiltonian can be diagonalised by Bogoliubov transformation

$$\begin{pmatrix} a_{\mathbf{k}} \\ a_{-\mathbf{k}}^\dagger \end{pmatrix} = \begin{pmatrix} u_0(\mathbf{k}) & v_0(\mathbf{k}) \\ v_0(\mathbf{k}) & u_0(\mathbf{k}) \end{pmatrix} \begin{pmatrix} \alpha_{\mathbf{k}} \\ \alpha_{-\mathbf{k}}^\dagger \end{pmatrix} \ , \tag{S31}$$

with coefficients

$$u_0(\mathbf{q}) = -\sqrt{\frac{A_0(\mathbf{q}) + \omega_0(\mathbf{q})}{2\omega_0(\mathbf{q})}} \ , \tag{S32a}$$

$$v_0(\mathbf{q}) = \sqrt{\frac{A_0(\mathbf{q}) - \omega_0(\mathbf{q})}{2\omega_0(\mathbf{q})}} \ . \tag{S32b}$$

to give

$$\mathcal{H}_{\text{LSWT}} = E_0 + \Delta E_0 + \sum_{\mathbf{q}} \omega_0(\mathbf{q}) \alpha_{\mathbf{q}}^\dagger \alpha_{\mathbf{q}} + \dots \ , \tag{S33}$$

where the magnon dispersion is given by

$$\omega_0(\mathbf{q}) = \sqrt{A_0(\mathbf{q})^2 - B_0(\mathbf{q})^2} \ , \tag{S34}$$

and the correction to the ground state energy coming from zero–point fluctuations is

$$\Delta E_0 = \sum_{\mathbf{q}} \left[ \omega_0(\mathbf{q}) - A_0(\mathbf{q}) \right] \ . \tag{S35}$$

This result forms the basis for the LSWT predictions shown in the main text.

We note that the coefficients of the Bogoliubov transformation used to diagonalise the Hamiltonian determine all ground–state averages, and will be important in determining interactions corrections, below.

### B. Mean field decoupling of interactions at $\mathcal{O}(1/S)$

To assess the effect of interactions we must return to the terms which are quartic in Bose operators. These have two sources, the products of interactions on different bonds which arise through the 4–spin interaction ($K$) [Eq. (S17)], and the terms $F_{ij}$ and $G_{ij}$ which introduce quartic interactions on an individual bond [Eq. (S24)]. At the lowest level of approximation, commonly known as "$1/S$ corrections", we can construct at new quadratic theory in which quantum fluctuations are treated at a Hartree–like mean–field level, and renormalize the dispersion of magnons, without causing them to decay into other states.

To do this, we we decouple quartic terms through pairwise contractions

$$ABCD = \langle AB \rangle CD + \langle AC \rangle BD + \langle AD \rangle BC + \langle BC \rangle AD + \langle BD \rangle AC + \langle CD \rangle AB \ . \tag{S36}$$

The resulting theory is once again quadratic in Bose operators, but depends on expectation values of pairs of operators

$$n \equiv \langle a_i^\dagger a_i \rangle \ , \tag{S37a}$$

$$t_n \equiv \langle a_i^\dagger a_{i+\delta_n} \rangle \ , \tag{S37b}$$

$$\Delta_n \equiv \langle a_i a_{i+\delta_n} \rangle \ , \tag{S37c}$$

where $\delta_n$ connects the lattice site $i$ to its $n^{\text{th}}$–neighbour. In physical terms, these new fields reflect the average of fluctuations on a single site ($n$), the transport of magnons along a bond ($t$), and the pairing of magnons on different sublattices ($\Delta$). These averages can be calculated within the LSWT defined in Section III A, to yield

$$\Delta_n = -\frac{1}{N} \sum_{\mathbf{q}} \gamma_n(\mathbf{q}) u_0(\mathbf{q}) v_0(\mathbf{q}) \ , \tag{S38a}$$

$$t_n = \frac{1}{N} \sum_{\mathbf{q}} \gamma_n(\mathbf{q}) v_0(\mathbf{q})^2 \ , \tag{S38b}$$

$$n = \frac{1}{N} \sum_{\mathbf{q}} v_0(\mathbf{q})^2 \ , \tag{S38c}$$

where $u_0(\mathbf{q})$ and $v_0(\mathbf{q})$ are defined through Eq. (S32). The results are presented in Fig. S8.

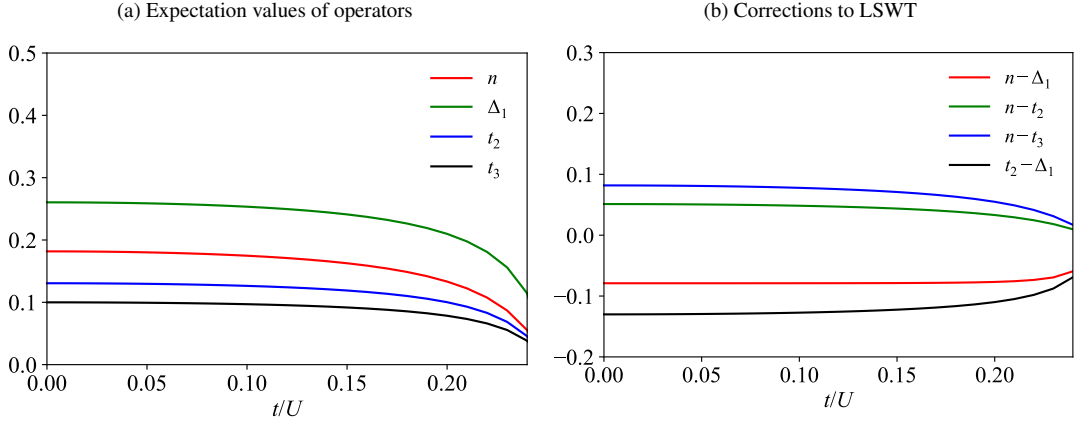

Figure S8. Coefficients entering $1/S$ corrections to linear spin wave theory (LSWT). (a) Expectation values of operators $n$, $\Delta_1$, $t_2$, $t_3$ [Eq. (S37)]. (b) Coefficients used in calculating corrections to LSWT [Eq. (S40)]. All coefficients are plotted as a function of $t/U$.

From this starting point we can proceed directly to a quadratic Hamiltonian of the same form of LSWT, Eq. (S28), at the expense some lengthly but straightforward algebra. However some economy of effort, and additional insight, can be gained by recasting problem in terms of an effective model containing only two-spin interactions. We accomplish this below.

### C. Analysis of dispersion including $1/S$ corrections

At the level a LSWT with $1/S$ corrections, the 4–spin terms in $\mathcal{H}_\sigma$ [S17] decouple into effective interactions on $1^{\mathrm{st}}$– and $2^{\mathrm{nd}}$–neighbour bonds. And for this reason, at this level of approximation, it is possible to recast $\mathcal{H}_\sigma$ as an effective $J_1$–$J_2$–$J_3$ Heisenberg model,

$$\mathcal{H}^{\mathrm{eff}} = J_1^{\mathrm{eff}} \sum_{\langle ij \rangle_1} \mathbf{S}_i \cdot \mathbf{S}_j + J_2^{\mathrm{eff}} \sum_{\langle ij \rangle_2} \mathbf{S}_i \cdot \mathbf{S}_j + J_3^{\mathrm{eff}} \sum_{\langle ij \rangle_3} \mathbf{S}_i \cdot \mathbf{S}_j \,, \tag{S39}$$

with exchange interactions

$$J_1^{\mathrm{eff}} = (J_1 - 2K) \left[ 1 - \frac{n}{S} + \frac{\Delta_1}{S} \right] + 4K \left[ \frac{n}{S} + \frac{t_2}{S} - \frac{2\Delta_1}{S} \right] \sim \mathcal{O}\left( \frac{t^2}{U} \right) \,, \tag{S40a}$$

$$J_2^{\mathrm{eff}} = (J_2 - K) \left[ 1 - \frac{n}{S} + \frac{t_2}{S} \right] + 2K \left[ \frac{n}{S} + \frac{t_2}{S} - \frac{2\Delta_1}{S} \right] \sim \mathcal{O}\left( \frac{t^4}{U^3} \right) \,, \tag{S40b}$$

$$J_3^{\mathrm{eff}} = J_3 \left[ 1 - \frac{n}{S} + \frac{t_3}{S} \right] \sim \mathcal{O}\left( \frac{t^4}{U^3} \right) \,. \tag{S40c}$$

where $\Delta_1$, $t_2$ and $t_3$ are defined through Eq. (S38). By construction, these effective interactions contain both the "bare interactions" found in the LSWT [Section III A], and the corrections arising at $1/S$, through the "mean field" terms $\Delta_1$, $t_2$ and $t_3$ [Section III C].

In Fig. S9 we show the effective exchange interactions $J_1^{\mathrm{eff}}$, $J_2^{\mathrm{eff}}$ and $J_3^{\mathrm{eff}}$ evolve with $t/U$. In the limit $t/U \to 0$ we recover a square–lattice Heisenberg AF

$$\lim \frac{t}{U} \to 0 : \qquad J_1^{\mathrm{eff}} \to Z_c J \quad , \quad J_2^{\mathrm{eff}} \to 0 \quad , \quad J_3^{\mathrm{eff}} \to 0 \,, \tag{S41}$$

where

$$Z_c \approx 1.158 \quad , \quad J = \frac{4t^2}{U} \,, \tag{S42}$$

consistent with published results for LSWT with $1/S$ corrections [24–31]. As $t/U$ is increased, $J^{\mathrm{eff}}$ is renormalized downwards, and competes with an emerging FM interaction on $2^{\mathrm{nd}}$–neighbour bonds. For physically relevant values of $t/U$, $J_3^{\mathrm{eff}}$ remains small, and has a negligible effect on dispersion.

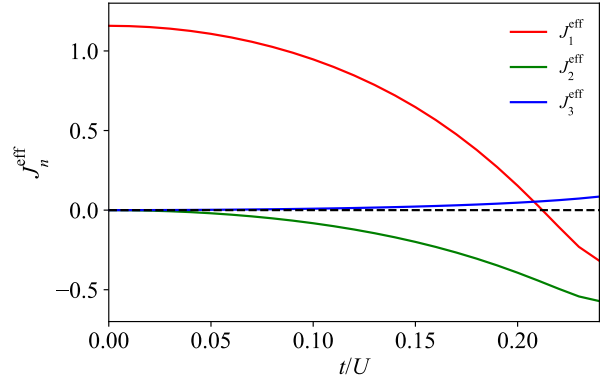

Figure S9. Effective exchange interactions $J_1^{\text{eff}}$, $J_2^{\text{eff}}$, $J_3^{\text{eff}}$ as a function of $t/U$, following Eq. (S40). Interactions are plotted relative to $J = 4t^2/U$ [Eq. (S42)]. The dominant terms are an effective antiferromagnetic exchange on $1^{\text{st}}$–neighbour bonds, $J_1^{\text{eff}} > 0$, and effective ferromagnetic exchange on $2^{\text{nd}}$ neighbour bonds, $J_2^{\text{eff}} < 0$. For $t/U \approx 0.212$, $J_1^{\text{eff}}$ changes sign, and interactions are no longer consistent with a $(\pi, \pi)$ Néel ground state.

Starting from $\mathcal{H}^{\text{eff}}$ [Eq. (S39)], we can now construct a theory which take account of leading interaction effects, simply by carrying out LSWT. Following Eq. (S24), we write

$$\mathcal{H}_{\text{LSWT}}^{\text{eff}} = E_0' + S\big[J_1^{\text{eff}} \sum_{\langle ij \rangle_1} f_{ij} + J_2^{\text{eff}} \sum_{\langle ij \rangle_2} g_{ij} + J_3^{\text{eff}} \sum_{\langle ij \rangle_3} g_{ij}\big] + \mathcal{O}\left(\frac{E_0'}{S^2}\right),$$ (S43)

and introduce Holstein–Primakoff Bosons as before [Eq. (S20)]. After Fourier transformation, the Hamiltonian takes the form Eq. (S28), with

$$A(\mathbf{q}) = 4SJ_1^{\text{eff}} + 4SJ_2^{\text{eff}}[\gamma_2(\mathbf{q}) - 1] + 4SJ_3^{\text{eff}}[\gamma_3(\mathbf{q}) - 1],$$ (S44a)

$$B(\mathbf{q}) = -4SJ_1^{\text{eff}}\gamma_1(\mathbf{q}).$$ (S44b)

The model can be diagonalized by a standard Bogoliubov transformation, yielding a magnon dispersion

$$\omega(\mathbf{q}) = \sqrt{A(\mathbf{q})^2 - B(\mathbf{q})^2}.$$ (S45)

This theory forms the basis of the results for interacting spin wave theory (LSWT + $1/S$) shown in the main text.

### D. Application to experiment

Comparison with experiment is most easily made through the dynamical structure factor $S_\perp(\mathbf{q}, \omega)$. To compute this, we employ the Lehmann representation

$$S_\perp(\mathbf{q}, \omega) = \sum_{n, \alpha} \delta(\omega - \omega(\mathbf{q})) |\langle n | S_{\mathbf{q}}^\alpha | 0 \rangle|^2, \quad \alpha = x, y,$$ (S46)

where $\omega(\mathbf{q})$ is defined through Eq. (S45). The higher order terms in $\mathcal{O}(1/S)$ that introduce magnon-magnon interactions into the Hamiltonian also introduce additional magnon correlation functions in the dynamical structure factor. Generically, these can modify the transverse contributions and introduce new longitudinal and transverse-longitudinal contributions, see e.g. [33]. For our model the transverse part is only modified by an unimportant overall scale factor $(1 - n/S)$ where $n$ is the magnon density (as the on-site term $\langle a_i^2 \rangle$ is zero). The transverse-longitudinal parts vanish due to the remnant $U(1)$ symmetry (which forbids one to two magnon decay). Finally, while the longitudinal parts are non-zero, as they are only sensitive to the two-magnon excitations; as the one-magnon excitations are the focus of this work we will not pursue these further. We thus see that the dynamical structure factor at $\mathcal{O}(1/S^2)$ can be computed using the usual linear spin-wave formalism, with interactions taken into account via the renormalized exchange constants.

Therefore, using Eq. (S20) and the Bogoliubov transformation, we obtain

$$S_\perp(\mathbf{q}, \omega) = \frac{A(\mathbf{q}) - B(\mathbf{q})}{4\omega(\mathbf{q})} \delta(\omega - \omega(\mathbf{q})),$$ (S47)

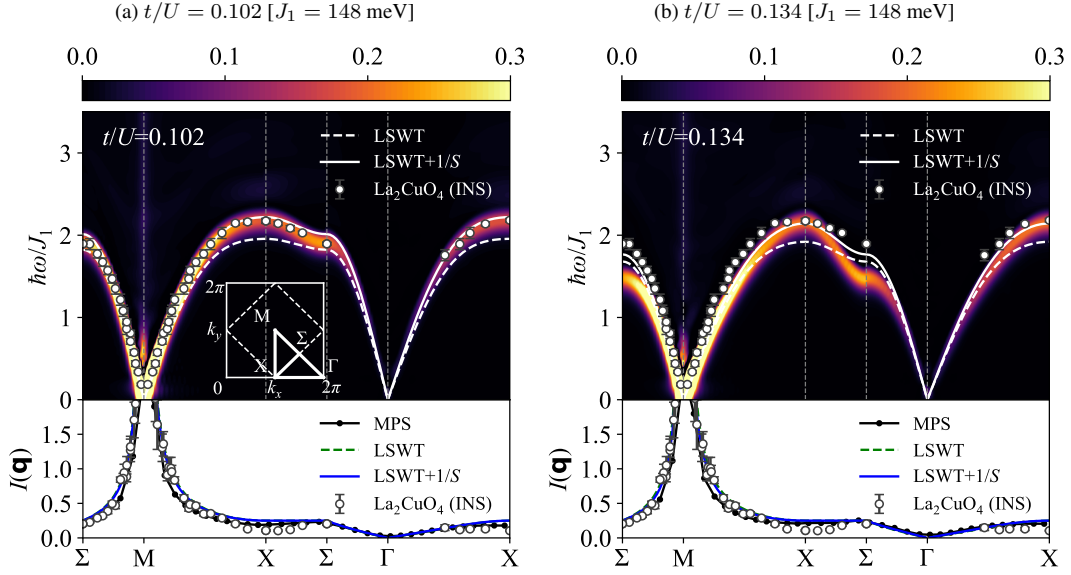

Figure S10. Comparison of excitations measured in La$_2$CuO$_4$ to theoretical predictions for different ratios of $t/U$. (a) Results for $t/U = 0.102$, as shown in main text. Predictions of linear spin wave theory (LSWT), and spin wave theory with leading $1/S$ corrections (LSWT + $1/S$) are overlaid on matrix product state (MPS) results for $S_\perp(\mathbf{q}, \omega)$ and $I_\perp(\mathbf{q})$. (b) Results for $t/U = 0.134$, the ratio proposed in [32], showing how quantum effects strongly renormalise the magnon dispersion. Attempts extract parameters from fits to low-order spin wave theory lead to systematic overestimates of $t/U$. All results were obtained for the model $\mathcal{H}_\sigma$ [Eq. (S7)], with parameters Eq. (S8), as described in the text.

where $A(\mathbf{q})$, $B(\mathbf{q})$ are defined through Eq. (S44b), and $\omega(\mathbf{q})$ through Eq. (S45). This result provides the basis for the results for dynamical structure factors described as "LSWT + $1/S$" in the main text. Examples of the predictions which follow from this theory are shown in Fig. S10.

### E. Effective two–parameter model

Given that $J_3^{\mathrm{eff}} \ll J_1^{\mathrm{eff}}, J_2^{\mathrm{eff}}$ for all relevant values of $t/U$ [Fig. S9], $\mathcal{H}^{\mathrm{eff}}$ [Eq. (S39)] can be further reduced to an effective 2–parameter model

$$\mathcal{H}_{J_1-J_2}^{\mathrm{eff}} = J_1^{\mathrm{eff}} \sum_{\langle ij \rangle_1} \mathbf{S}_i \cdot \mathbf{S}_j + J_2^{\mathrm{eff}} \sum_{\langle ij \rangle_2} \mathbf{S}_i \cdot \mathbf{S}_j \tag{S48}$$

where antiferromagnetic (AF) $1^{\mathrm{st}}$–neighbour interaction,

$$J_1^{\mathrm{eff}} \sim \mathcal{O}\left(\frac{t^2}{U}\right) > 0 \tag{S49}$$

competes with ferromagnetic (FM) $2^{\mathrm{nd}}$–neighbour interaction,

$$J_2^{\mathrm{eff}} \sim \mathcal{O}\left(\frac{t^4}{U^3}\right) < 0 \tag{S50}$$

The success of low-order spin wave theory in describing the magnon dispersion in La$_2$CuO$_4$ [32, 34] can ultimately be traced to the fact that the magnon dispersion is relatively well–described the two–parameter model Eq. (S48) [4]. Within this approach the dominant AF interaction $J_1^{\mathrm{eff}}$ is used to set the overall scale of magnon dispersion, while the sub–dominant FM interactions $J_2^{\mathrm{eff}}$ controls the dispersion on the magnetic Brillouin Zone boundary (which vanishes for $J_2 = 0$).

In the context of a one–band Hubbard model where [Eq. S40], this means that it is possible to use $J_1$ to fix the overall magnon bandwidth, and then tune $J_2$ by varying $t/U$, such that the dispersion fits experiment. However because LSWT dramatically underestimates quantum effects, this leads to systematic overestimates of $t/U$, as shown in Fig. 1(d) of the main text. Including $1/S$ corrections leads to better estimates, but the parameters found are still systematically too high.

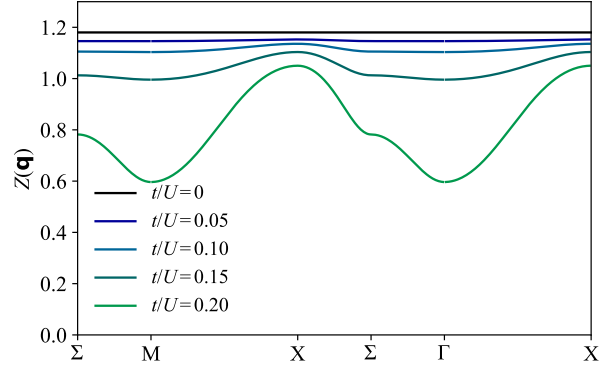

Figure S11. Momentum dependent rescaling factor $Z(\mathbf{q})$ [Eq. (S51)] found in leading $1/S$ correction to linear spin wave theory (LSWT), for a range of values of $t/U$. In the limit $t/U \to 0$, the model $H_\sigma$ [Eq. (S7)] reduces to a nearest–neighbour Heisenberg antiferromagnet, and $1/S^2$ corrections lead to a uniform renormalization of dispersion $Z(\mathbf{q}) = Z_c = 1.18$ [24–26, 28]. For all finite $t/U$, the form of dispersion is modified by $1/S$ corrections. These corrections are largely governed by the finite value of $J_c/J_1$ [Eq. (S8)].

## F. Comment on form of corrections to dispersion

A common practice in the fitting of experimental data [32, 34] has been to multiply the dispersion obtained in linear spin wave theory (LSWT) for $\mathcal{H}_\sigma$ by a constant renormalization factor $Z_c = 1.18$, coming from calculations of $1/S^2$ corrections for a nearest–neighbour Heisenberg model [24–26, 28], cf. Fig. S9. As the results above show, this approximation is justified only in the limit $t/U \to 0$, where $1/S$ corrections do not change the form of spin wave dispersion.

To quantify the error which arises when data is analyzed in this way at finite $t/U$, we compute the momentum–dependent rescaling factor

$$Z(\mathbf{q}) = \frac{\omega(\mathbf{q})}{\omega_0(\mathbf{q})}. \tag{S51}$$

In Fig. S11 we show the result of $Z(\mathbf{q})$ in high–symmetry directions, for a range of values of $t/U$. The trend is for increasing deviation from the dispersion predicted by LSWT with increasing $t/U$, with a progressive reduction in spin wave velocity, and a softening of excitations at $\mathbf{q}_\Sigma = (\pi/2, \pi/2)$ relative to those at $\mathbf{q}_X = (\pi, 0)$. These effects are the qualitatively similar to those seen in MPS calculations. However the magnitude of corrections to dispersion is smaller within LSWT + $1/S$, leading to systematic overestimates of $t/U$ when compared with experimental data.

In the specific case of $La_2CuO_4$, this leads to estimates of $t/U \approx 0.134$ [32, 34]. The comparison of MPS and LSWT ($+1/S$) calculations with experiment at this value is presented in Fig. S10. In this case, $1.0 \lesssim Z(\mathbf{q}) \lesssim 1.1$ and the error in dispersion coming from the approximation $Z(\mathbf{q}) = Z_c = 1.18$ is of order 10%.

This approximation can be relaxed by working directly with LSWT + $1/S$ theory developed in Section III C and Section III D, leading to the results shown in the main text.



\end{document}